\documentclass[12pt]{article}
\usepackage{hyperref}
\usepackage{epsfig}
\usepackage{latexsym}

\newskip\defaultbaselineskip\defaultbaselineskip=12pt

\def\GeV{\mathord{\rm \;GeV}}
\def\MeV{\mathord{\rm \;MeV}}

\def\bar{\overline}

\def\gsim{\mathrel{\raise2pt\hbox to 8pt{\raise -5pt\hbox{$\sim$}\hss{$>$}}}}
\def\rsim{\mathrel{\raise2pt\hbox to 8pt{\raise -5pt\hbox{$\sim$}\hss{$>$}}}}
\def\lsim{\mathrel{\raise2pt\hbox to 8pt{\raise -5pt\hbox{$\sim$}\hss{$<$}}}}
\def\ssqr#1#2{{\vbox{\hrule height.#2pt
      \hbox{\vrule width.#2pt height#1pt \kern#1pt\vrule width.#2pt}
      \hrule height.#2pt}\kern-.#2pt}}

\def\Dsl{\,\raise.15ex\hbox{$/$}\mkern-13.5mu D} 
\def\dsl{\raise.15ex\hbox{$/$}\kern-.57em\hbox{$\partial$}}

    \def\CS{{\cal S}} 
    \def\CW{{\cal W}}

\ifx\href\undefined\def\href#1#2{{#2}}\fi
\def\spireshome{http://www.slac.stanford.edu/cgi-bin/spiface/find/hep/www?FORMAT=WWW&}
{\catcode`\%=12
\xdef\spiresjournal#1#2#3{\noexpand\protect\noexpand\href{\spireshome
                          rawcmd=find+journal+#1%2C+#2%2C+#3}}
\xdef\spireseprint#1#2{\noexpand\protect\noexpand\href{\spireshome rawcmd=find+eprint+#1%2F#2}}
\xdef\spiresreport#1{\noexpand\protect\noexpand\href{\spireshome rawcmd=find+rept+#1}}
\xdef\spireskey#1{\noexpand\protect\noexpand\href{\spireshome key=#1}}
}
\def\eprint#1#2{\spireseprint{#1}{#2}{#1/#2}}
\def\report#1{\spiresreport{#1}{#1}}
\def\nohref{}

\def\putpaper{\edef\refpage{\the\count0}%
              \def\nohref{}%
              {\def\ {+}\def\nohref##1{}\edef\temp{\noexpand\spiresjournal
               {\journalname}{\volume}{\refpage}}\expandafter}\temp
               {\sfcode`\.=1000{\journalname} {\bf \volume} (\refyear)
                \refpage}\egroup}
\def\putpage{\edef\refpage{\the\count0}%
              \def\nohref{}%
              {\def\ {+}\def\nohref##1{}\edef\temp{\noexpand\spiresjournal
               {\journalname}{\volume}{\refpage}}\expandafter}\temp
              {\refpage}\egroup}
\def\dojournal#1#2 (#3){\def\journalname{#1}\def\volume{#2}\def\refyear
                        {#3}\afterassignment\putpaper\bgroup\count0=}
\def\morepage{\afterassignment\putpage\bgroup\count0=}

\def\NPB#1{\dojournal{Nucl.\ Phys.}{B#1}}
\def\NPBPS#1{\dojournal{Nucl.\ Phys.\ \nohref(Proc.\ Suppl.\nohref)}{\nohref B#1}}

\def\PRL#1{\dojournal{Phys.\ Rev.\ Lett.}{#1}}
\def\PR#1{\dojournal{Phys.\ Rev.}{#1}}
\def\PRep#1{\dojournal{Phys.\ Rep.}{#1}}

\def\PRD#1{\dojournal{Phys.\ Rev.}{D#1}}

\def\PLB#1{\dojournal{Phys.\ Lett.}{#1B}}

\def\ZPC#1{\dojournal{Z.\ Phys.}{C#1}}

\def\etal{{\it et al.}}

\textheight=9.5in
\textwidth=7.0in
\hoffset=-0.85in
\voffset=-1.20in

\def\gsim{\mathrel{\raise2pt\hbox to 8pt{\raise -5pt\hbox{$\sim$}\hss{$>$}}}}
\def\rsim{\mathrel{\raise2pt\hbox to 8pt{\raise -5pt\hbox{$\sim$}\hss{$>$}}}}
\def\lsim{\mathrel{\raise2pt\hbox to 8pt{\raise -5pt\hbox{$\sim$}\hss{$<$}}}}

\newcommand\figcaption[1]{\vskip-0.0truein\caption{#1}\vskip0.0truein}
\newcommand\alphamsbar{\hbox{$\alpha_{\overline{MS}}$}}

\newcommand\MSbar{{\hbox{$\overline{MS}$}}}
\newcommand\mmsbar{\hbox{$m_{\overline{MS}}$}}
\newcommand\mbar{\hbox{$\overline{m}$}}
\def\cpt{\hbox{$ \chi $PT}}
\def\trace{{\rm tr}}
\def\vev#1{\langle #1 \rangle}

\setlength\hfuzz{3pt}
\hbadness=5000

\begin{document}

\begin{titlepage}
 \null
 \begin{center}
 \makebox[\textwidth][r]{LAUR-96-1840}
 \par\vspace{0.25in} 
  {\Large
      Light Quark Masses from Lattice QCD}
  \par
 \vskip 2.0em
 
 {\large 
  \begin{tabular}[t]{c}
        Rajan Gupta\footnotemark\ and
          Tanmoy Bhattacharya\footnotemark\\[0.5em]
        \em Group T-8, Mail Stop B-285, Los Alamos National Laboratory\\
        \em Los Alamos, NM 87545, U.~S.~A\\[1.5em]
  \end{tabular}}
 \par \vskip 2.0em
 {\large\bf Abstract}
\end{center}

\quotation We present estimates of the masses of light quarks using
lattice data.  Our main results are based on a global analysis of all
the published data for Wilson, Sheikholeslami-Wohlert (clover), and
staggered fermions, both in the quenched approximation and with
$n_f=2$ dynamical flavors.  We find that the values of masses with the
various formulations agree after extrapolation to the continuum limit
for the $n_f=0$ theory.  Our best estimates, in the \MSbar\ scheme at
$\mu=2 \GeV$, are $\mbar=3.4 \pm 0.4 \pm 0.3 \MeV$ and $m_s = 100 \pm
21 \pm 10 \MeV$ in the quenched approximation. The $n_f=2$ results,
$\mbar = 2.7 \pm 0.3 \pm 0.3 \MeV$ and $m_s = 68 \pm 12 \pm 7 \MeV$,
are preliminary.  (A linear extrapolation in $n_f$ would further
reduce these estimates for the physical case of three dynamical
flavors.)  These estimates are smaller than phenomenological estimates
based on sum rules, but maintain the ratios predicted by chiral
perturbation theory (\cpt).  The new results have a significant impact
on the extraction of $\epsilon'/\epsilon$ from the Standard Model.
Using the same lattice data we estimate the quark condensate using the
Gell-Mann-Oakes-Renner relation. Again the three formulations give
consistent results after extrapolation to $a=0$, and the value turns
out to be correspondingly larger, roughly preserving $m_s \vev{\bar
\psi \psi}$.

{
 \def\test{Hfootnote.2}
 \expandafter\ifx\csname @currentHref\endcsname\test
 \expandafter\def\csname @currentHref\endcsname{Hfootnote.1}
 \fi
 \footnotetext[1]{Email: rajan@qcd.lanl.gov}
}
\footnotetext{Email:    tanmoy@qcd.lanl.gov}
\vfill
\mbox{15 April, 1997}
\end{titlepage}

\setlength{\textfloatsep}{12pt plus 2pt minus 2pt}

\makeatletter 

\setlength{\leftmargini}{\parindent}
\def\@listi{\leftmargin\leftmargini
            \topsep 0\p@ plus2\p@ minus2\p@\parsep 0\p@ plus\p@ minus\p@
            \itemsep \parsep}
\long\def\@maketablecaption#1#2{#1. #2\par}

\advance \parskip by 0pt plus 1pt minus 0pt

\makeatother

\section{INTRODUCTION}
\label{s_intro}

The masses of light quarks $m_u$, $m_d$, and $m_s$ are three of the
least well known parameters of the Standard Model. The range of values
listed in the particle data book \cite{rPDB94}, $2 \le m_u \le 8$, $5
\le m_d \le 15$, and $100 \le m_s \le 300$ (evaluated in the \MSbar\ scheme at
$\mu=1$ GeV), is indicative of the large uncertainty. These quark
masses have to be inferred from the masses of low lying hadrons.
Theoretically, the best defined procedure is chiral perturbation
theory (\cpt) which relates the masses of pseudoscalar mesons to
$m_u,\ m_d$, and $m_s$.  Recall that the lowest order chiral
Lagrangian is
\begin{equation}
{\cal L} = {f^2 \over 8}  \trace \bigg[\big( 
          \partial_\mu \Sigma \partial_\mu \Sigma^\dagger \big) 
       + 2 \mu \big(M\Sigma + M\Sigma^\dagger) \bigg] 
\label{eq:cptL_0}
\end{equation}
The presence of the overall unknown scale $\mu$ implies that only
ratios of quark masses can be determined using \cpt.  The predictions
from \cpt\ for the two independent ratios are 
\cite{gasserPR, rMq90leutwyler, rMq96leutwyler} \\
\medskip

\begin{tabular}{ccc}
                          &  Lowest order      & Next order \\
$         2  m_s / (m_u + m_d) $   &  $25    $ & $  24.4(1.5)  $ \phantom{\ .} \\ \\
$            m_u / m_d         $   &  $0.55  $ & $  0.553(43)  $ . \\
\label{tab:mqcpt}
\end{tabular}
\medskip

\noindent 
One notes that the change from lowest to next order is
insignificant. Unfortunately, the prospects for further improvement
within \cpt\ are not good.  To get absolute numbers, one has to
combine the \cpt\ analysis with sum rule or with some other model
calculations.  The two most recent and up-to-date versions of the sum
rule analysis has been performed by Bijnens, Prades, and de Rafael
\cite{BPR95} giving $m_u + m_d = 12(2.5) \MeV$ and by Jamin and Munz
\cite{Jamin95} giving $m_s = 178(18) \MeV$. It is typical of sum rule
analysis to quote results at $\mu=1$ GeV, while lattice results are
presented at $\mu=2$ GeV. To facilitate comparison, we translate the
sum-rule values to $\mu=2$ GeV, whereby $m_u + m_d = 9.4(1.8) \MeV$
and $m_s = 126(13) \MeV$ using $\Lambda_{\overline{MS}}^{(3)} = 300$
and $380$ MeV respectively.  These values are about one $\sigma$ above
the quenched lattice results we present below.  However, as discussed
in \cite{BGM96}, a reanalysis of the sum-rules calculations show that
the uncertainties are large enough to preclude any serious
disagreement with the lattice results.  To improve upon the sum-rule
estimates requires hard to get at experimental information on the
hadronic spectral function. Thus, we believe that lattice QCD offers
the best approach to determining these quantities, and this paper
presents an analysis of the current data.

In lattice QCD the quark masses are input parameters in the
simulations.  Their values are determined by tuning the masses of an
equal number of hadrons to their physical values. One additional
hadron mass is needed to fix the lattice scale, or equivalently the
value of $\alpha_s$.  In our lattice simulations we do not include
electromagnetic effects, so we can only calculate the isospin
symmetric light quark mass $\mbar=(m_u+m_d)/2$. The hadrons we use to
fix the lattice scale are $M_\rho= 770 \MeV$ (one could instead, for
example, use $f_\pi= 131 \MeV$, but its determination is less reliable
and has not been reported for all the data we use), while to fix
$\mbar$ and $m_s$ we study the behavior of pseudoscalar and vector
meson masses as a function of the quark mass.  (For experimental
numbers we use the isospin averaged values $M_\pi= 137 \MeV$, $M_K=495
\MeV$, $M_\phi= 1020 \MeV$ and $M_K^* = 894 \MeV$.)
We have chosen to use these pseudoscalar and vector
mesons since the corresponding correlation functions have been
measured with the smallest statistical errors.  Thus, in principal,
quark masses can be determined without any ambiguity from these
calculations.

Current lattice simulations make a number of approximations because of
which estimates have systematic errors in addition to statistical
errors.  Some of these issues have been discussed in the reviews by
Ukawa \cite{mq92ukawa} and Gupta \cite{mq94gupta}.  In this paper we
present an analysis of the cumulative world data as available in
November 1996 for staggered, Wilson, and Sheikholeslami-Wohlert
($c_{SW}=1$) fermions, and both for the quenched and $n_f=2$ theories.
(The parameter $c_{SW}$ is the strength of the clover term that is
added to the Wilson action to remove the $O(a)$ discretization errors
\cite{SWaction, luscher96imp}.)  We present evidence showing that the largest
uncertainty in the results arises from discretization ($O(a)$) errors
and from the dependence on the number of dynamical flavors. We find
that the discretization errors show the expected behavior, $O(a)$ for
Wilson and $O(a^2)$ for staggered. For clover action one expects
corrections of $O(\alpha_s a)$, however the current data show behavior
similar to that with Wilson fermions.  By combining the results from
these three formulations we show that the discretization errors can be
controlled.  Thus the main remaining uncertainty is due to
extrapolation in $n_f$, because of which we are only able to extract
rough estimates of the quark masses for the physical case of $n_f=3$.
These are significantly smaller than the phenomenological estimates
mentioned above, and lead to a large enhancement in the standard model
prediction of $\epsilon'/\epsilon$.

Lastly, we also present an analysis of the quark condensate,
$\vev{\bar \psi \psi}$, using the data for pseudoscalar meson mass and
the Gell-Mann-Oakes-Renner relation.  Again we find that the three
lattice discretizations of the Dirac action give consistent results
after extrapolation to the continuum limit. The condensate turns out
to be roughly a factor of two larger than phenomenological estimates
such that the renormalization group invariant quantity $m \vev{\bar
\psi \psi}$ is preserved.

This paper is organized as follows. In Section \ref{s_massdef} we
summarize the definition of the quark mass, the relation between the
lattice and continuum results, and the 2-loop running. The
reorganization of the lattice perturbation theory $a\ la$ Lepage and
Mackenzie is discussed in Section \ref{s_Zfactors}. A brief
description of the lattice data used in this analysis is given in
Section \ref{s_tech}. Analyses of $\mbar$ and $m_s$ are presented in
Sections~\ref{s_mbar} and \ref{s_ms}. We discuss the variation of
quark masses with the clover coefficient in Sheikholeslami-Wohlert
action in Section~\ref{s_clover}. A general discussion of the systematic
errors is given in Section~\ref{s_syserror}. In Section~\ref{s_comparison} we
compare our results with earlier calculations and discuss the impact
on $\epsilon'/\epsilon$ in Section~\ref{s_epsilon}. The calculation of
the quark condensate is presented in Sections~\ref{s_xxgmor}. We end
with our conclusions and future outlook in
Section~\ref{s_conclusions}.

\section{DEFINITION OF QUARK MASSES}
\label{s_massdef}

There are two ways of calculating the \MSbar\ mass for light quarks at
scale $\mu$ from lattice estimates. The first is
\begin{eqnarray}
\mmsbar(\mu) \ &=& \ Z_m(\mu a) m_L(a) \nonumber \\
               &=& \bigg\{ 1 - \lambda \big[ 8 log(\mu a) - C_m \big] \bigg\} m_L(a) \nonumber \\
               &\approx& \bigg\{1 - \lambda \big[ 8 log(\mu a) - (C_m -tad) \big] \bigg\} 
                    \ \big\{m_L(a)  X \big\}
\label{eq:connection}
\end{eqnarray}
where $Z_m \equiv 1/Z_S$ is the mass renormalization constant relating
the lattice and the continuum regularization schemes at scale $\mu$,
and $\lambda = g^2/16\pi^2$.  The 1-loop perturbative expressions for
$Z_m$ are given in Table~\ref{t1_zfac}.  For Wilson (or Sheikholeslami-Wohlert (SW) clover)
type of fermions, the lattice estimate of the quark mass, defined at
scale $q^*$, is taken to be $m_L(q^*) a = ({1 / 2\kappa} - {1/ 2
\kappa_c} )$.  We have also carried out the analysis using the
alternate definition ${\rm Ln}(1+({1/2\kappa}-{1/2\kappa_c}))$. The
change in individual estimates is $\le 3\%$ for data at $a \lsim 0.5
\GeV^{-1}$ and negligible in the extrapolated values.  Since the
analysis with this $O(a)$ improved definition is much less transparent
we use the simpler form.  For staggered fermions $m_L(q^*) = m_0$, the
input mass.  The last form in Eq.~\ref{eq:connection} shows the result
of applying tadpole improvement as discussed in
Section~\ref{s_Zfactors}.

The second method is to use the Ward Identity for the renormalized
axial vector current, $\partial_\mu Z_A A_\mu = (m_1 + m_2) Z_P P +
O(a)$, where $m_1$ and $m_2$ are the masses of the two fields in the
bilinear operators. The quark mass, to $O(a)$, is then given by the
following ratio of correlation functions
\begin{equation}
(m_1+m_2) = {Z_A \over Z_P}   {\langle \partial_4 A_4(\tau) J(0) \rangle \over 
                                        \langle P(\tau) J(0) \rangle } .
\label{eq:wardidentity1}
\end{equation}
where $J(0)$ is any interpolating field operator that produces pions
at $\vec p = 0$. Here, $Z_A$ and $ Z_P $ are the renormalization
constants for the axial and pseudoscalar densities that match the
lattice and continuum theories at scale $\mu = q^*$. The important
point to note is that since the Ward identity is valid locally, the
value of $m$ should be independent of $\tau$. Violations of this
relation are signals for $O(a)$ errors in the currents. This has been
exploited by the Alpha collaboration to remove the $O(a)$
discretization errors non-perturbatively in the clover action and
currents \cite{luscher96imp}.

At long time separation ($\tau$ large) one can assume that the 
2-point correlation function is described by the asymptotic form $\sim
e^{-m_\pi \tau}$. One can then write Eq.~\ref{eq:wardidentity1} as
\begin{equation}
(m_1+m_2) = {Z_A \over Z_P} {m_\pi}  {\langle A_4(\tau) J(0) \rangle \over 
                                        \langle P(\tau) J(0) \rangle } .
\label{eq:wardidentity2}
\end{equation}
There are very few calculations of $m_q$ using these two versions of
the Ward Identity. Also, at this stage the perturbative and
non-perturbative estimate of $Z_P$ do not agree \cite{Allton94}. For
these reasons we postpone the discussion of this method to a later work
\cite{WI96LANL}.

Once \mmsbar\ has been calculated at the continuum scale $\mu$, its value at any 
other scale $Q$ is given by the two loop running \cite{Allton94}
\begin{equation}
{m(Q) \over m(\mu)}  = \bigg({g^2(Q) \over g^2(\mu)}\bigg)^{\gamma_0 / 2 \beta_0} \ 
        \bigg( 1 + {g^2(Q) - g^2(\mu) \over 16\pi^2} \big( 
        {\gamma_1 \beta_0 - \gamma_0 \beta_1 \over 2 \beta_0^2} \big) \bigg) . 
\label{eq:mrunning}
\end{equation}
where in the \MSbar\ scheme for $N$ colors and $n_f$ flavors 
\begin{eqnarray}
\beta_0 &=&  {11N - 2n_f \over 3}  , \nonumber \\
\beta_1 &=&  {34 \over 3} N^2 - {10 \over 3} Nn_f - {N^2-1 \over N} n_f , \nonumber \\
\gamma_0 &=& 6 {N^2-1 \over 2N}  ,  \nonumber \\
\gamma_1 &=& {97N \over 3}{N^2-1 \over 2N} + 3( {N^2-1 \over 2N})^2 
             - {10n_f \over 3}{N^2-1 \over 2N} \ . 
\label{eq:NDRvalues}
\end{eqnarray}
We will quote our final numbers at $Q=2 \GeV$ as we do not feel
confident using the 2-loop relations to run to smaller scales.  

\section{Renormalization Constants}
\label{s_Zfactors}

Our final results are stated in the \MSbar\ scheme.  The required
1-loop renormalization constants for matching the lattice and
continuum operators for Wilson, clover, and staggered versions of the lattice
discretization are collected in Table~\ref{t1_zfac} using results derived in
\cite{MartiZhang, smitMg, Gockeler}.  We reorganize lattice perturbation theory 
using the Lepage-Mackenzie prescription \cite{Lepage}. This
prescription involves four parts: the renormalization of the quark
field $Z_\psi$ and the quark mass, the removal of tadpole contribution from the 1-loop
operator renormalization, the choice of $\alpha_s$, and an
estimate of the typical scale $q^*$ characterizing the lattice calculation. 

\begin{table} 
\caption{Renormalization constants in the \MSbar\ scheme before
tadpole subtraction for staggered~\protect\cite{smitMg}, and Wilson and clover ($c_{SW}=1$)
\protect\cite{Gockeler} fermions. Here $\lambda = g^2 /16\pi^2$. \looseness=-1}
\def\q{\quad}
\def\s{\phantom{-}}
\def\sq{\phantom{-}\q}
\def\NDR{{NDR} }
\def\dredMS{DRED}
\def\mua{{\rm ln}(\mu a) }
\def\lam{\lambda}
\newcommand\0{hphantom{0}}
\setlength{\tabcolsep}{3pt}
$$
\begin{tabular}{|l|c|c|c|}
\hline
$        $&$ Staggered                  $&$ Wilson                    $&$ Clover(c_{SW}=1)         $\cr
\hline								                                     
$ Z_A    $&$ 1                          $&$ 1 - 21.06 \lam           $&$ 1 - 18.39 \lam             $\cr
$ Z_P    $&$ 1/Z_m                      $&$ 1 + \lam (8\mua - 30.13) $&$ 1 + \lam (8\mua - 29.84) $\cr
$ Z_m    $&$ 1 - \lam (8\mua - 52.288)  $&$ 1 - \lam (8\mua - 17.27) $&$ 1 - \lam (8\mua - 25.75) $\cr
\hline
\end{tabular}
$$

\label{t1_zfac}
\end{table}

We use the following Lepage-Mackenzie definition of the strong coupling 
constant \cite{Lepage, HEMCGCw, HEMCGC94hmks}, 
\begin{eqnarray}
-{\rm ln} \langle {1 \over 3 }{\rm Tr} plaq \rangle &= &
              {4\pi \over 3} \alpha_V(3.41/a) \ (1 - (1.191+0.025n_f) \alpha_V) \qquad (Wilson), \nonumber \\
	&=&   {4\pi \over 3} \alpha_V(3.41/a) \ (1 - (1.191+0.070n_f) \alpha_V) \qquad (Stag.) , 
\label{eq:alphavdef}
\end{eqnarray}
from which the \MSbar\ coupling at scale $3.41/a$ is given by 
\begin{equation}
\alphamsbar(3.41/a) = \alpha_V(e^{5/6}3.41/a) ( 1 + {2 \over \pi} \alpha_V) .
\label{eq:alphamsdef}
\end{equation}
The value of \alphamsbar\ at any other scale is then obtained by
integrating the standard 2-loop $\beta$-function for the appropriate
number of flavors. The results are given in Tables~\ref{t_Qlist},
\ref{t_Dlist}, and \ref{t_Clist}. Ideally, $\alphamsbar(2\GeV)$ should
be independent of $\beta$.  From these tables one can see the
variation with $\beta$, and the extent to which the variation in $1/a$
at fixed $\beta$ feeds into $\alphamsbar(2\GeV)$.  Reorganizing the
lattice perturbation theory in terms of $\alphamsbar$ has the advantage that
the continuum and lattice $\alpha_s$ is the same when matching the
theories at scale $\mu = q^*$.  We call this matching procedure
``horizontal'' matching \cite{BKW96LANL}.

An estimate of $q^*$ is not straightforward since $Z_m$ is
logarithmically divergent.  So we appeal to the general philosophy of
tadpole improvement, $i.e.$ it is designed to remove the
short-distance lattice artifacts.  Once these have been removed, the
typical scale $q^*$ of the lattice calculation becomes less
ultraviolet.  Thus, our preferred scheme is one in which $q^*=\mu=
1/a$, which we call $TAD1$.  To analyze the dependence of the results
on $q^*$ we also investigate the choices $q^*=\mu= 0.5/a, 2/a$ and $
\pi/a$.  A more detailed description of these schemes and our
implementation of the matching between the lattice and continuum is
given in \cite{BKW96LANL}.

In all methods of calculating the quark mass,
Eqs.~\ref{eq:connection}, \ref{eq:wardidentity1},
\ref{eq:wardidentity2}, the normalization of the quark fields does not
enter, or cancels between the different currents.  Thus, one does not
have to consider this factor.  The quark mass gets scaled by the tadpole 
factor $X$, $i.e.$ $m_L \to X m_L$, where we use the non-perturbatively 
determined estimate for $X$. 

The tadpole factor $X$ for Wilson and clover fermions is
chosen to be $8 \kappa_c$ \cite{MFF93LANL}. It has the perturbative expansion 
$X = 1 + tad \ \lambda$ where 
\begin{eqnarray}
tad  &=& 17.14  \ , \qquad (Wilson) \ , \nonumber \\
tad  &=& 10.66  \ , \qquad (Clover\ (c_{SW}=1)) \ .
\label{eq:defU0Wilson}
\end{eqnarray}
For Wilson fermions at, say, $\beta=6.0$, $\lambda = 0.0153$ in the
$TAD1$ prescription, so the perturbative value $8\kappa_c=1.262$ is
almost identical to our non-perturbative estimate $8\kappa_c = 1.257$
\cite{rHM96LANL}.  The Lepage-Mackenzie reorganization, as shown in
Eq.~\ref{eq:connection}, therefore seems benign, nevertheless, we
include tadpole subtraction as its purpose, in general, is to
reorganize lattice perturbation theory to make the neglected higher
order corrections small.  The entries in Table~\ref{t1_zfac} show that
after tadpole subtraction the 1-loop corrections are indeed smaller
for Wilson and clover fermions.  However, note that the 1-loop results
presented in \cite{Gockeler}, and used here, show that the
perturbative correction grows with $c_{SW}$.

For staggered fermions we define the tadpole factor $X$ as the inverse
of the fourth root of the expectation value of the plaquette
\begin{equation}
X = U_0^{-1} = plaquette^{-1/4} = (1 + 13.16 \lambda ) \ .
\label{eq:defU0stag}
\end{equation}
The agreement, at $\beta=6.0$, between the perturbative value $1.20$
and the non-perturbative value $1.14$ is still reasonable in $TAD1$,
even though the plaquette is designed to match at $q^* \approx 3.41/a$
as shown in Eq.~\ref{eq:alphavdef}.  The point to note is that for any
reasonable choice of $X$, the correction $C_m - tad$ is still very large
for staggered fermions .  Thus one might doubt whether the 1-loop
perturbative result for $Z_m$ is reliable. It turns out, as we show
later, that using the expressions given in Table~\ref{t1_zfac} give
results that, in the $a=0$ limit, agree with those obtained using
Wilson-like fermions.

The second noteworthy outcome of our analysis is that the dependence
of the results on the choice of $q^*$, and whether or not one does
tadpole subtraction, are small.  We find that $m$ increases with
$q^*$, and the most significant variation is in the quenched staggered
data ($\sim 5\%$). The difference between tadpole subtraction and no
tadpole subtraction is significant only at strong coupling ($\sim 3\%$
at $\beta=6.0$ for the quenched theory, and $< 3\%$ for $\beta \ge
5.4$ for $n_f=2$), and even smaller for different choices of tadpole
factor.  Based on the above estimates we assume that for $a \lsim 0.5$
the uncertainty in relating the lattice results to those in a
continuum scheme is under control at the $5\%$ level.  We consider
this variation negligible since for many of the data points it is
small compared to even the statistical errors.  Furthermore, this
uncertainty is one aspect of the discretization errors, and our
estimate of the errors associated with the extrapolation to $a\to 0$
incorporates it. To summarize, all of our lattice estimates are quoted
using the TAD1 scheme, and the uncertainty associated with this choice
is included in our quoted extrapolation error.


\section{LATTICE PARAMETERS OF DATA ANALYZED}
\label{s_tech}

The list of calculations from which we have taken data are summarized
in Tables~\ref{t_Qlist} and \ref{t_Dlist}. The quenched Wilson data
are taken from Refs.~\cite{HEMCGCw, rHM96LANL, APEw, HM96JLQCD,
GF11.spectrum, QCDPAXw, APE96, UKQCDw}, quenched staggered from
\cite{HEMCGC94hmks, APEks, KS96JLQCD, USstaghm91, DKSstag93,
OHTAstag96}, quenched clover from \cite{APE96}, dynamical ($n_f=2$)
Wilson from \cite{USdynW2,HEMCGC94W2, dbetaW296SCRI, SESAMdynW2}, and dynamical
($n_f=2$) staggered fermions from \cite{HEMCGC94hmks, Tsukuba94hmks2,
Columbia91hmks2}.  All these calculations use the simple Wilson
(plaquette) action for the gauge fields.  Thus the difference between
staggered and Wilson-like fermion data for fixed $n_f$ is a reflection
of the difference in $O(a)$ errors between the various discretization
schemes.

\begin{table} 
\caption{Lattice parameters of the quenched data used in the global
analysis. We list the reference, the type of fermion action ($W$=Wilson,
$S$=staggered, $C$=clover with $c_{SW}=1$), the coupling
$\beta=6/g^2$, the lattice size, the number of configurations in the
statistical sample, the values of quark masses in lattice units
($\kappa$ values) used in the fits for staggered (Wilson and clover)
fermions, the lattice scale $1/a \GeV$ extracted from
$M_\rho$, and the values of $\alpha_\MSbar$ at $q^* = 1/a$ and $2\GeV$. }
\def\q{\quad}
\def\s{\phantom{-}}
\newcommand\0{hphantom{0}}
\setlength{\tabcolsep}{2pt}
\begin{tabular}{|l|c|c|c|c|l|c|cc|}
\hline
      &      &         & Lattice & \# of &                           & Scale $1/a$  &%
          \multispan{2}{\hfil$\alpha_{\MSbar}(q^*)$\hfil\vrule}\\
Label & Ref. & $\beta$ & size    & Conf. & Quark masses used in the fits & (GeV)         &%
          $(1/a)$&$(2\GeV)$\\
\hline
W1  & \cite{APEw}            & 5.7 &$24^3\times 32$& 50 & 0.165,  0.167,  0.168          &$1.431( 27)$&$0.246$&$0.211$\\
W2  & \cite{GF11.spectrum}   & 5.7 &$24^3\times 32$& 92 & 0.165,  0.1663, 0.1675         &$1.422( 24)$&$0.246$&$0.210$\\
W3  & \cite{QCDPAXw}         & 5.85&$24^3\times 54$& 100& 0.1585, 0.1595, 0.1605         &$1.958(114)$&$0.213$&$0.211$\\
W4  & \cite{HEMCGCw}         & 5.85&$16^3\times 32$& 90 & 0.1585, 0.1600                 &$1.741( 76)$&$0.213$&$0.201$\\
W5  & \cite{HM96JLQCD}       & 5.9 &$16^3\times 40$& 150& 0.156,  0.157,  0.158,  0.1585 &$1.987( 48)$&$0.205$&$0.204$\\
W6  & \cite{GF11.spectrum}   & 5.93&$24^3\times 36$& 210& 0.156,  0.1573, 0.1581         &$2.000( 37)$&$0.201$&$0.201$\\
W7  & \cite{HEMCGCw}         & 5.95&$16^3\times 32$& 90 & 0.1554, 0.1567                 &$1.941( 58)$&$0.198$&$0.196$\\
W8  & \cite{rHM96LANL}       & 6.0 &$32^3\times 64$& 170& 0.155,  0.1558, 0.1563         &$2.338( 43)$&$0.192$&$0.205$\\
W9  & \cite{APEw}            & 6.0 &$24^3\times 32$& 78 & 0.155,  0.1558, 0.1563         &$2.204( 70)$&$0.192$&$0.200$\\
W10 & \cite{QCDPAXw}         & 6.0 &$24^3\times 54$& 200& 0.155,  0.1555, 0.1563         &$2.423(146)$&$0.192$&$0.208$\\
W11 & \cite{APE96}           & 6.0 &$18^3\times 64$& 320& 0.153,  0.154,  0.155          &$2.154( 64)$&$0.192$&$0.198$\\
W12 & \cite{HM96JLQCD}       & 6.1 &$24^3\times 64$& 100& 0.152,  0.153,  0.154,  0.1543 &$2.629( 59)$&$0.181$&$0.201$\\
W13 & \cite{GF11.spectrum}   & 6.17&$32^3\times 40$& 219& 0.1519, 0.1526, 0.1532         &$2.755( 48)$&$0.176$&$0.198$\\
W14 & \cite{UKQCDw}          & 6.2 &$24^3\times 48$&  18& 0.1523, 0.1526, 0.1529         &$2.735(172)$&$0.173$&$0.194$\\
W15 & \cite{APE96}           & 6.2 &$24^3\times 64$& 250& 0.1510, 0.1515, 0.1520, 0.1526 &$2.914( 89)$&$0.173$&$0.199$\\
W16 & \cite{APE96}           & 6.2 &$24^3\times 64$& 110& 0.1510, 0.1520, 0.1526         &$2.934(121)$&$0.173$&$0.200$\\
W17 & \cite{HM96JLQCD}       & 6.3 &$32^3\times 80$& 100& 0.150,  0.1505, 0.151,  0.1513 &$3.260( 88)$&$0.166$&$0.197$\\
W18 & \cite{APEw}            & 6.3 &$24^3\times 32$& 128& 0.1485, 0.1498, 0.1505         &$3.092( 68)$&$0.166$&$0.193$\\
W19 & \cite{APEw}            & 6.4 &$24^3\times 60$&  15& 0.1485, 0.1490, 0.1495         &$3.628(416)$&$0.159$&$0.195$\\
W20 & \cite{APE96}           & 6.4 &$24^3\times 64$& 400& 0.1488, 0.1492, 0.1496, 0.1500 &$4.095(163)$&$0.159$&$0.205$\\
\hline				   
S1  & \cite{APEks}           & 5.7 &$24^3\times 32$& 50 & 0.02, 0.015, 0.01, 0.005       &            &       &       \\
    & \cite{USstaghm91}      & 5.7 &$16^3\times 32$& 32 & 0.02, 0.015, 0.01, 0.005       &$0.951( 78)$&$0.246$&$0.180$\\
S2  & \cite{HEMCGC94hmks}    & 5.85&$16^3\times 32$& 160& 0.025, 0.01                    &$1.312( 75)$&$0.213$&$0.180$\\
S3  & \cite{KS96JLQCD}       & 5.85&$16^3\times 32$&  60& 0.01,  0.02, 0.03, 0.04        &$1.340( 31)$&$0.213$&$0.182$\\
S4  & \cite{KS96JLQCD}       & 5.93&$20^3\times 40$&  50& 0.01,  0.02, 0.03, 0.04        &$1.571( 33)$&$0.201$&$0.183$\\
S5  & \cite{HEMCGC94hmks}    & 5.95&$16^3\times 32$& 190& 0.025, 0.01                    &$1.645( 25)$&$0.198$&$0.184$\\
S6  & \cite{APEks}           & 6.0 &$24^3\times 32$& 60 & 0.04, 0.02, 0.01               &$1.843( 78)$&$0.192$&$0.187$\\
S7  & \cite{USstaghm91}      & 6.0 &$24^3\times 40$& 23 & 0.03, 0.025, 0.02, 0.015, 0.01 &$1.911( 84)$&$0.192$&$0.189$\\
S8  & \cite{DKSstag93}       & 6.0 &$32^3\times 64$& 200& 0.01, 0.005, 0.0025            &$1.917( 34)$&$0.192$&$0.189$\\
S9  & \cite{KS96JLQCD}       & 6.0 &$24^3\times 64$&  50& 0.02,  0.03, 0.04              &$1.855( 41)$&$0.192$&$0.187$\\
S10 & \cite{USstaghm91}      & 6.2 &$32^3\times 48$& 23 & 0.025, 0.015, 0.01, 0.005      &$2.569( 78)$&$0.173$&$0.189$\\
S11 & \cite{KS96JLQCD}       & 6.2 &$32^3\times 64$&  40& 0.005, 0.01, 0.02              &$2.616( 91)$&$0.173$&$0.191$\\
S12 & \cite{USstaghm91}      & 6.4 &$32^3\times 48$& 24 & 0.015, 0.010, 0.005            &$3.467(376)$&$0.159$&$0.192$\\
S13 & \cite{KS96JLQCD}       & 6.4 &$40^3\times 96$& 40 & 0.005, 0.01,  0.02             &$3.429( 78)$&$0.159$&$0.191$\\
S14 & \cite{DKSstag93}       & 6.5 &$32^3\times 64$& 100& 0.01, 0.005, 0.0025            &$3.962(127)$&$0.152$&$0.192$\\
S15 & \cite{OHTAstag96}      & 6.5 &$48^3\times 64$& 200& 0.01, 0.005, 0.0025            &$3.811( 59)$&$0.152$&$0.189$\\
\hline				   
C1  & \cite{APE96}           & 6.0 &$18^3\times 64$& 200& 0.1425, 0.1432, 0.1440             &$1.867( 81)$&$0.192$&$0.187$\\
C2  & \cite{APE96}           & 6.2 &$24^3\times 64$& 250& 0.14144, 0.14184, 0.14224, 0.14264 &$2.604(131)$&$0.173$&$0.190$\\
C3  & \cite{APE96}           & 6.2 &$18^3\times 64$& 200& 0.14144, 0.14190, 0.14244          &$3.082(384)$&$0.173$&$0.204$\\
C4  & \cite{APE96}           & 6.4 &$24^3\times 64$& 400& 0.1400, 0.1403, 0.1406, 0.1409     &$3.951(182)$&$0.159$&$0.202$\\
\hline
\end{tabular}

\label{t_Qlist}
\end{table}

\begin{table} 
\caption{Lattice parameters of the $n_f=2$ dynamical simulation data we use in the global
analysis. The rest is same as in Table~\protect\ref{t_Qlist}. \looseness=-1}
\vskip 12pt
\def\q{\quad}
\def\s{\phantom{-}}
\newcommand\0{hphantom{0}}
\setlength{\tabcolsep}{2pt}
\begin{tabular}{|l|c|c|c|c|l|c|cc|}
\hline
      &      &         & Lattice & \# of &                           & Scale $1/a$  &%
          \multispan{2}{\hfil$\alpha_{\MSbar}(q^*)$\hfil\vrule}\\
Label & Ref. & $\beta$ & size    & Conf. & Quark masses used in the fits & (GeV)         &%
          $(1/a)$&$(2\GeV)$\\
\hline
$\CW_2 1$& \cite{dbetaW296SCRI}   & 5.3 &$16^3\times 32$&417-484& 0.167, 0.1675         &$1.890(110)$&$0.266$&$0.259$\\
$\CW_2 2$& \cite{USdynW2}         & 5.4 &$16^3\times 32$& 14-15 & 0.160, 0.161,  0.162  &$1.467(113)$&$0.254$&$0.223$\\
$\CW_2 3$& \cite{USdynW2}         & 5.5 &$16^3\times 32$& 15-27 & 0.158, 0.159,  0.160  &$1.812( 97)$&$0.227$&$0.219$\\
$\CW_2 4$& \cite{dbetaW296SCRI}   & 5.5 &$16^3\times 32$&400-669& 0.1596,0.1600, 0.1604 &$2.179( 84)$&$0.226$&$0.234$\\
$\CW_2 5$& \cite{USdynW2}         & 5.6 &$16^3\times 32$& 32-45 & 0.156, 0.157          &$2.332(209)$&$0.213$&$0.226$\\
$\CW_2 6$& \cite{SESAMdynW2}      & 5.6 &$16^3\times 32$&  100  & 0.156, 0.157,  0.1575 &$2.379( 77)$&$0.213$&$0.227$\\
\hline				    
$\CS_2 1$& \cite{HEMCGC94hmks}    & 5.6 &$16^3\times 32$& 200-400 & 0.025, 0.01              &$1.798( 73)$&$0.232$&$0.223$\\
$\CS_2 2$& \cite{Columbia91hmks2} & 5.7 &$16^3\times 32$&$\sim$50 & 0.025, 0.02, 0.015, 0.01 &$2.341( 78)$&$0.216$&$0.230$\\
$\CS_2 3$& \cite{Tsukuba94hmks2}  & 5.7 &$20^3\times 20$& 150-160 & 0.02,  0.01              &$2.236( 99)$&$0.216$&$0.226$\\
\hline
\end{tabular}

\label{t_Dlist}
\end{table}

In principle, if lattice simulations could be done with three
dynamical flavors and with realistic parameters, one could tune the
three quark masses to reproduce the spectrum and thereby determine
their values. However, since this is not the case, the strategy we use
is as follows. For each value of the lattice parameters ($\beta$,
$n_f$, fermion action) we fit the data for $M_\pi^2$ and $M_\rho$ as a
linear function of the quark mass
\begin{eqnarray}
M_\pi^2 &=& A_\pi  + B_\pi  (m_1 + m_2)/2  \nonumber \\
M_\rho  &=& A_\rho + B_\rho (m_1 + m_2)/2  \ ,
\label{e:chiralfits}
\end{eqnarray}
where $m_1$ and $m_2$ are the masses of the two quarks.  (We use
$M_\pi$ as shorthand for pseudoscalars, and $M_\rho$ for vector
mesons.)  These fits are made assuming that the data at each value of
the quark mass are independent, and the statistical error estimates
quoted by the authors are used in the $\chi^2$ minimization procedure.
Note that for Wilson-like fermions we make fits to $1/2\kappa$ whereby
$A_\pi$ defines $\kappa_c$ and $A_\rho$ is the intercept at
$1/2\kappa=0$. For staggered fermions $A_\pi$ should be zero due to
chiral symmetry. However, in our analysis we leave it as a free
parameter and use it to define the zero of the quark mass.  This
leaves $B_\pi, A_\rho, B_\rho$ from which we determine the three
quantities that we are interested in; the scale using $M_\rho$,
$\mbar$ using $M_\pi^2/M_\rho^2$, and $m_s$ in three different ways
using $M_K, \ M_{K^*},\ M_\phi$. Throughout the analysis we assume
that $\phi$ is a pure $s \bar s$ state.

The error analysis of the global data is somewhat limited since in
most cases we only have access to the final published numbers. For
this reason we cannot include two types of correlations in the data
when making fits using Eq.~\ref{e:chiralfits}. Firstly, those between
pseudoscalar and vector mesons at a given quark mass, and secondly
those between the meson masses at different quark masses calculated on
the same background gauge configurations as is the case for the
quenched simulations. Thus, when making chiral fits we simply assume
that the errors in the lattice measurements of meson masses are
uncorrelated.  Thereafter, the errors in $A_\pi, B_\pi, A_\rho,
B_\rho$ are propagated self-consistently to the final estimates. By
comparing results in a few cases where a full error analysis has been
done, we find that these shortcomings change estimates of individual
points by less than even the statistical errors.  The one exception is
the results from QCDPAX collaboration \cite{QCDPAXw} where the
neglected correlations have a large effect. They quote
significantly different values for $A_\pi, B_\pi, A_\rho, B_\rho$
based on correlated fits. Using their correlated fit parameters give quark masses
consistent with other estimates at the same couplings.  Thus, we
believe that our estimates based on a reanalysis of the global data
are reliable.

The input parameters and results ($\alpha_\MSbar$, lattice scale
$1/a$, $A_\pi, B_\pi, A_\rho, B_\rho$, and the quark masses in lattice
units and in $\MeV$ in $\MSbar$ scheme at $2\ \GeV$)
for each data set are given in Tables~\ref{t_Qlist}, \ref{t_Dlist},
\ref{t_Clist}, \ref{t_mqresultsQ}, \ref{t_mqresultsD}, and
\ref{t_mqresultsC}.  (In the case of data from
Ref.~\cite{KS96JLQCD}, the analysis started with the results
for $A_\pi, B_\pi, A_\rho, B_\rho$ because the raw data have not been
published.)  We note that the various data points are obtained on
lattices of different physical volumes, the statistical sample size
varies significantly (the sample size is rather small in some cases),
and the strategy for extracting masses varies from group to
group. However, it turns out that in the calculation of quark masses,
there is a cancellation of errors that make these differences much
less significant than for example in the estimates of meson masses themselves.
We substantiate some of these remarks by presenting a detailed
analysis of some of the systematic errors in Section \ref{s_syserror}
using data from our high statistics calculation presented in
Ref.~\cite{rHM96LANL}.  

Once we have the quark masses at different lattice scales, we analyze
the dependence on the lattice spacing $a$ by comparing Wilson, clover,
and staggered results.  Where data permit, we omit points at the
stronger couplings (larger $a$) for the following two reasons. First,
we use only the leading correction in the extrapolation to $a=0$, and
secondly, the perturbative matching becomes less reliable as $\beta$
is decreased.  The end result is that we find that the leading
corrections give a good fit to the data, and in the $a =0$ limit the
different fermion formulations give consistent results.

To investigate the $n_f$ dependence, we compare the $n_f = 0$ and
$n_f=2$ results. We then assume a linear behavior in $n_f$ to 
extrapolate these two points to estimate the desired physical
value. As we make clear in our discussions, the extrapolation in $n_f$
turns out to be the weakest part of our analysis.

\section{LIGHT QUARK MASS $\mbar=(m_u+m_d)/2$}
\label{s_mbar}

In order to extract light quark masses from lattice simulations we
have used the simplest ansatz, Eq.~\ref{e:chiralfits}, for the chiral
behavior of hadron masses.  The reason for this truncation is that in
most cases the data for $M_\pi$ and $M_\rho$ exist at only $2-4$
values of ``light'' quark masses in the range $0.3m_s - 2m_s$.  In
this restricted range of quark masses the existing data do not show
any significant deviation from linearity.  Using linear fits to
$M_\pi^2$ data over a limited range in $m_q$ means that we can predict
only one independent quark mass from the pseudoscalar data, and we
cannot test the difference between tree level and NLO results
predicted by $\cpt$ analyses.  The mass we prefer to extract using the
pseudoscalar spectrum, barring the complications of quenched \cpt\
which are expected to become significant only for $m_q \lsim 0.3 m_s$
\cite{mq94gupta, REV96SHARPE}, is $\mbar$ because to extract $m_s$
using $m_K$ needs first an extrapolation to $\mbar$ in the light
quark.  While this choice avoids the question whether lowest order
\cpt\ is valid up to $m_s$, one must bear in mind that the fits are
made to data in the ``heavier'' range $0.3m_s \le m_q \sim 2 m_s$ and
then extrapolated to $\mbar$.  The bottom line is that to get $\mbar$
we linearly extrapolate the ratio $M_\pi^2/M_\rho^2$ in $m$ to its
physical value $0.03166$.

\begin{table} 
\caption{Results for the fit parameters $\kappa_c$ for Wilson-like and
$A_\pi$ for staggered fermions, $B_\pi$, $A_\rho$ and $B_\rho$,
defined in Eq.~\protect\ref{e:chiralfits}, and the quark masses for
the quenched simulations. Both the lattice values, $\mbar a$ and $m_s
(M_\phi) a$, as well as the physical values, $\mbar(\MSbar, 2 \GeV)$
and $m_s(\MSbar, 2 \GeV, M_\phi)$ in $\MeV$, are given.  The lattice
parameters and the references to the original work are given in
Table~\protect\ref{t_Qlist}. Chiral fits for labels marked by a ${}^*$ 
have $\chi^2/dof \ge 3.0$. }
\vskip -18pt
$$
\setlength{\tabcolsep}{1pt}
\begin{tabular}{|l|cccc|cc|cc|}
 \hline
 &&&&&\multispan2{\hfil$\mbar$\hfil\vrule}&\multispan2{\hfil$m_s(M_\phi)$\hfil\vrule}\\
 Label&%
  $\kappa_c/A_\pi$&$B_\pi$&$A_\rho$&$B_\rho$&%
  $\bar m a \times 10^5$&$\bar m$&%
  ${m_s} a  \times 10^3$&$ {m_s}$\\
  \hline
$W 1$&$0.169250(    11)$&$  2.758(  43)$&$ -4.221( 481)$&$  1.609( 160)$&$332( 13)$&$ 6.05( 14)$&$112( 13)$&$ 204(  20)$\\
$W 2$&$0.169308(    12)$&$  2.712(  38)$&$ -4.223( 422)$&$  1.612( 140)$&$342( 12)$&$ 6.19( 13)$&$113( 11)$&$ 203(  17)$\\
$W 3$&$0.161605(    02)$&$  2.334(  33)$&$ -6.840(1372)$&$  2.336( 436)$&$210( 24)$&$ 5.30( 31)$&$ 57( 14)$&$ 143(  26)$\\
$W 4$&$0.161662(    17)$&$  2.317(  75)$&$ -4.161(1150)$&$  1.487( 366)$&$267( 24)$&$ 5.89( 30)$&$ 99( 28)$&$ 219(  53)$\\
$W 5$&$0.159811(    11)$&$  2.309(  39)$&$ -5.819( 471)$&$  1.982( 148)$&$206( 10)$&$ 5.24( 14)$&$ 66( 06)$&$ 167(  12)$\\
$W 6$&$0.158950(    06)$&$  2.141(  32)$&$ -5.086( 473)$&$  1.738( 148)$&$219( 08)$&$ 5.59( 13)$&$ 74( 07)$&$ 189(  16)$\\
$W 7$&$0.158379(    13)$&$  2.166(  49)$&$ -4.453( 701)$&$  1.535( 218)$&$230( 14)$&$ 5.64( 20)$&$ 86( 14)$&$ 211(  29)$\\
$W 8$&$0.157143(    03)$&$  1.993(  26)$&$ -6.663( 521)$&$  2.196( 162)$&$172( 06)$&$ 5.20( 11)$&$ 50( 05)$&$ 152(  11)$\\
$W 9$&$0.157122(    08)$&$  2.037(  60)$&$ -5.680( 993)$&$  1.894( 309)$&$190( 13)$&$ 5.35( 22)$&$ 62( 12)$&$ 174(  28)$\\
$W10$&$0.157135(    05)$&$  2.005(  50)$&$ -7.556(1561)$&$  2.473( 485)$&$159( 19)$&$ 5.02( 32)$&$ 43( 11)$&$ 136(  26)$\\
$W11$&$0.157024(    06)$&$  2.130(  24)$&$ -5.494( 477)$&$  1.837( 146)$&$190( 11)$&$ 5.21( 16)$&$ 65( 07)$&$ 179(  14)$\\
$W12$&$0.154978(    03)$&$  1.862(  20)$&$ -6.537( 400)$&$  2.116( 122)$&$146( 07)$&$ 4.97( 12)$&$ 46( 04)$&$ 158(   9)$\\
$W13$&$0.153760(    04)$&$  1.740(  31)$&$ -6.365( 477)$&$  2.042( 145)$&$142( 05)$&$ 5.06( 11)$&$ 46( 04)$&$ 163(  11)$\\
$W14$&$0.153308(    14)$&$  1.746( 104)$&$ -6.128(1603)$&$  1.964( 487)$&$144( 19)$&$ 5.06( 41)$&$ 48( 14)$&$ 169(  41)$\\
$W15$&$0.153292(    02)$&$  1.678(  17)$&$ -6.632( 604)$&$  2.113( 183)$&$132( 08)$&$ 4.99( 16)$&$ 42( 05)$&$ 159(  13)$\\
$W16$&$0.153292(    03)$&$  1.708(  24)$&$ -7.021( 793)$&$  2.232( 240)$&$128( 11)$&$ 4.87( 21)$&$ 39( 06)$&$ 151(  16)$\\
$W17$&$0.151796(    02)$&$  1.546(  22)$&$ -6.832( 631)$&$  2.145( 190)$&$114( 06)$&$ 4.86( 14)$&$ 37( 04)$&$ 157(  14)$\\
$W18$&$0.151637(    15)$&$  1.734(  31)$&$ -6.968( 290)$&$  2.188(  86)$&$113( 05)$&$ 4.53( 10)$&$ 38( 02)$&$ 152(   6)$\\
$W19$&$0.150632(   132)$&$  1.517( 236)$&$ -7.511(1686)$&$  2.326( 502)$&$ 94( 18)$&$ 4.47( 58)$&$ 31( 09)$&$ 145(  31)$\\
$W20$&$0.150580(    05)$&$  1.385(  31)$&$ -8.435( 696)$&$  2.596( 208)$&$ 81( 06)$&$ 4.42( 18)$&$ 24( 03)$&$ 133(  10)$\\
\hline					                		
$S 1^*$&$-0.0005429(  05)$&$  7.806(  33)$&$  0.791(  71)$&$  8.879(4405)$&$266( 43)$&$ 4.56( 37)$&$ 32( 17)$&$  54(  26)$\\
$S 2  $&$-0.0007702(  79)$&$  6.986(  49)$&$  0.583(  34)$&$  5.333(1392)$&$156( 18)$&$ 3.66( 21)$&$ 37( 12)$&$  86(  22)$\\
$S 3  $&$                $&$  6.700(  20)$&$  0.567(  13)$&$  4.960( 280)$&$156( 07)$&$ 3.74( 09)$&$ 39( 02)$&$  94(   5)$\\
$S 4  $&$                $&$  6.150(  10)$&$  0.484(  10)$&$  4.840( 200)$&$124( 05)$&$ 3.48( 07)$&$ 34( 02)$&$  96(   4)$\\
$S 5  $&$-0.0007099( 174)$&$  5.840(  47)$&$  0.466(  07)$&$  5.133( 377)$&$119( 04)$&$ 3.50( 06)$&$ 30( 03)$&$  89(   7)$\\
$S 6  $&$-0.0002067( 154)$&$  5.634(  45)$&$  0.414(  18)$&$  5.323( 588)$&$ 98( 08)$&$ 3.25( 14)$&$ 26( 04)$&$  87(  10)$\\
$S 7  $&$-0.0000768( 174)$&$  5.666(  65)$&$  0.398(  18)$&$  5.736( 786)$&$ 91( 08)$&$ 3.14( 14)$&$ 24( 04)$&$  82(  11)$\\
$S 8^*$&$-0.0001874(  03)$&$  5.690(  20)$&$  0.396(  08)$&$  7.717(1197)$&$ 90( 03)$&$ 3.11( 06)$&$ 18( 03)$&$  61(   9)$\\
$S 9  $&$                $&$  5.610(  10)$&$  0.410(  09)$&$  5.140( 170)$&$ 97( 04)$&$ 3.25( 07)$&$ 27( 01)$&$  91(   3)$\\
$S10^*$&$-0.0000717(  57)$&$  4.020(  29)$&$  0.297(  09)$&$  5.019( 484)$&$ 71( 04)$&$ 3.28( 10)$&$ 20( 02)$&$  93(   9)$\\
$S11  $&$                $&$  4.010(  20)$&$  0.291(  10)$&$  4.880( 400)$&$ 68( 05)$&$ 3.24( 11)$&$ 20( 02)$&$  96(   8)$\\
$S12  $&$ 0.0000310( 239)$&$  2.884( 110)$&$  0.219(  24)$&$  4.851(1850)$&$ 54( 12)$&$ 3.39( 38)$&$ 15( 07)$&$  97(  36)$\\
$S13  $&$                $&$  2.970(  20)$&$  0.222(  05)$&$  4.760( 200)$&$ 54( 02)$&$ 3.33( 08)$&$ 16( 01)$&$  98(   4)$\\
$S14  $&$-0.0001986( 107)$&$  2.482(  56)$&$  0.193(  06)$&$  4.880( 723)$&$ 48( 03)$&$ 3.45( 13)$&$ 13( 02)$&$  95(  14)$\\
$S15^*$&$ 0.0001403(  24)$&$  2.518(  18)$&$  0.199(  03)$&$  3.937( 365)$&$ 51( 02)$&$ 3.52( 06)$&$ 17( 02)$&$ 119(  11)$\\
\hline					                		
$C 1$&$0.145483(    03)$&$  2.674(  29)$&$ -6.265(1006)$&$  1.942( 288)$&$201( 18)$&$ 5.33( 24)$&$ 71( 13)$&$ 188(  27)$\\
$C 2$&$0.143150(    02)$&$  2.076(  27)$&$ -7.378(1373)$&$  2.196( 389)$&$133( 13)$&$ 4.99( 26)$&$ 45( 10)$&$ 169(  29)$\\
$C 3$&$0.143100(    08)$&$  2.110(  93)$&$-10.897(2879)$&$  3.189( 815)$&$ 94( 23)$&$ 4.26( 54)$&$ 26( 10)$&$ 120(  30)$\\
$C 4$&$0.141439(    04)$&$  1.695(  40)$&$-10.849( 949)$&$  3.123( 266)$&$ 71( 06)$&$ 4.16( 19)$&$ 21( 03)$&$ 123(  10)$\\
\hline
\end{tabular}
$$

\label{t_mqresultsQ}
\end{table}

\begin{figure}[t]
\hbox{\hskip15bp\epsfxsize=0.9\hsize\epsfbox{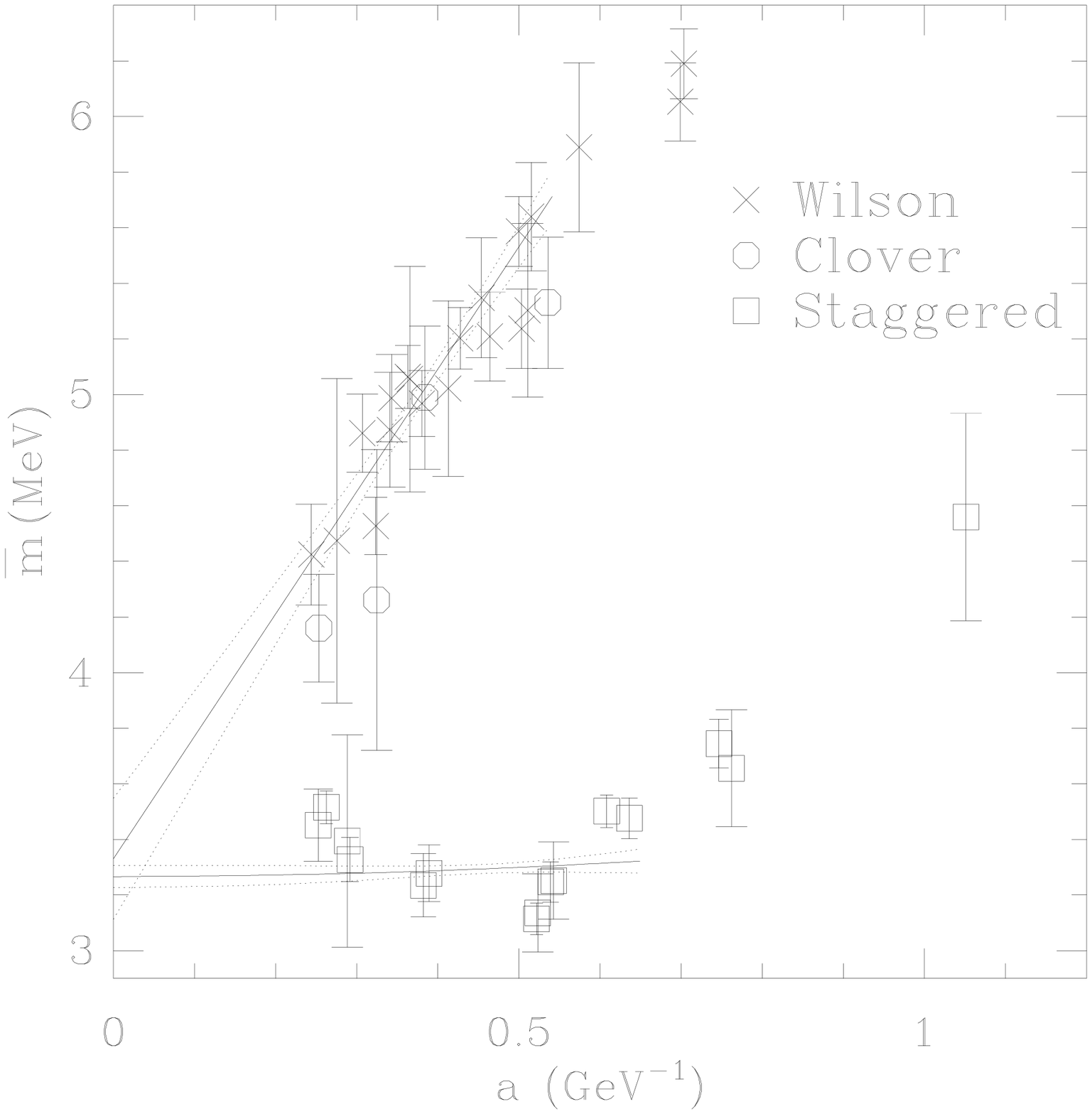}}
\vskip \baselineskip
\figcaption{The behavior of $\mbar(\MSbar,2 \GeV)$ extracted using the
quenched $M_\pi$ data in the TAD1 lattice scheme defined in the text.
The scale is set by $M_\rho$.  We do not show the fit to the clover
data for clarity.}  
\label{f_mbar}
\end{figure}

The global quenched data are shown in Fig.~\ref{f_mbar} and listed in
Table~\ref{t_mqresultsQ} for the tadpole subtraction scheme TAD1.  The
expected leading term in the discretization errors is $O(a)$ for
Wilson, $O(g^2 a)$ for clover, and $O(a^2)$ for staggered. The fits,
keeping only the leading dependence (for clover data we assume a
linear dependence in $a$), give
$$
\begin{tabular}{lll}
$m(a) = 3.33(22)\MeV\ [1 +  1.3(2) \GeV\ a]$ &  $\chi^2/dof=0.8$ & Wilson \\
$m(a) = 3.13(42)\MeV\ [1 +  1.4(4) \GeV\ a]$ &  $\chi^2/dof=0.6$ & Clover \\
$m(a) = 3.27(04)\MeV\ [1 + (0.20(11) \GeV\ a)^2]$ &  $\chi^2/dof=4.6$ & Staggered \,.
\label{tab:Qmlfita}
\end{tabular}
$$
For Wilson and staggered fermions we fit to $\beta \ge 5.93$ ($a \lsim
0.5 \GeV^{-1}$) to minimize the effect of higher order corrections.
The staggered data show a small rise for $\beta \ge 6.0$ after the
initial fall.  With present data it is not clear whether this is due
to poor statistics, the decreasing physical volume of the lattices
used, or higher order terms in $a$ and $g^2$. As a result, the fit
with only the lowest order correction is not able to capture this
trend in the data. This is reflected in the high $\chi^2/dof$. Thus we
consider, as an alternate estimate of the continuum result, the mean,
$\mbar = 3.4(1) \MeV$, of the estimates at the two highest $\beta$.  A
second alternative is to fit the data for $\beta \ge 6.0$. This give
$\mbar = 3.53(6)\ \MeV$ with a $\chi^2/dof = 0.95$. The corresponding
fit for Wilson fermions gives $\mbar = 3.45(23)\ \MeV$ with a
$\chi^2/dof = 0.9$.

\begin{figure}[t]
\hbox{\hskip15bp\epsfxsize=0.9\hsize\epsfbox{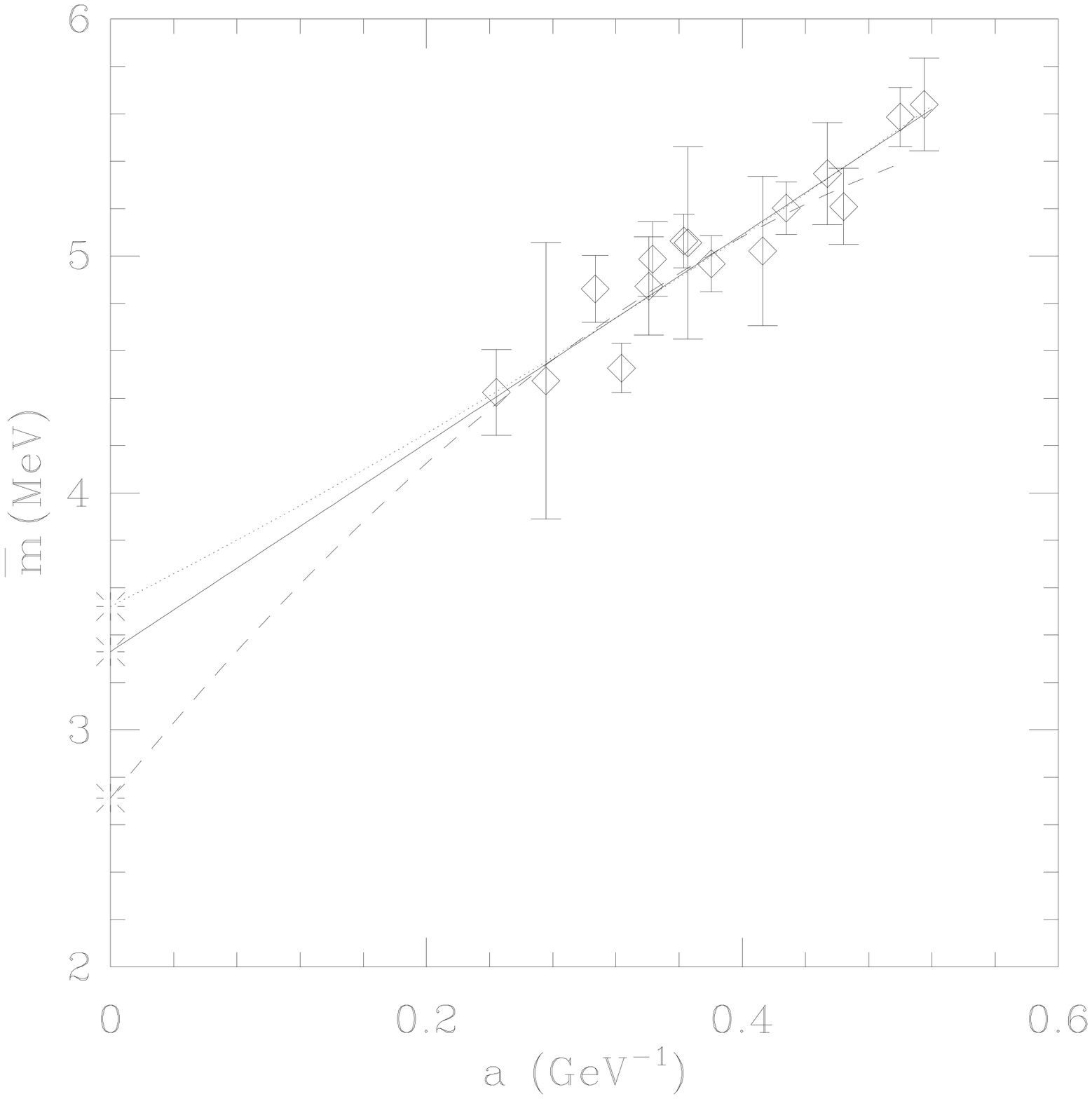}}
\vskip \baselineskip
\figcaption{Three different fits to quenched Wilson data for
$\mbar(\MSbar,2 \GeV)$.  The solid line is a linear fit to points at
$\beta \ge 5.93$, while the dotted line includes a quadratic
correction. The dashed line is the quadratic fit to points at $\beta
\ge 6.0$.}
\label{f_Wquad}
\end{figure}

Given the size of $O(a)$ corrections in the case of Wilson fermions,
we have also fit the data including a quadratic correction as shown in
Fig.~\ref{f_Wquad}. Unfortunately, we find that even the sign of the
quadratic term changes depending on whether we include all the points
with $\beta \ge 5.93$ or only those with $\beta \ge 6.0$. Also, one
cannot distinguish between the three fits on the basis of $\chi^2/dof$
which lies between $0.8-0.96$. The lack of stability of the
quadratic fits indicates that possible $O(a^2)$ corrections cannot be
determined with the current data.  For this reason we use the results
obtained using the lowest order fit as our best estimate.

The clover ($c_{SW}=1$) fermion results are surprisingly close to the
Wilson values even though the discretization errors should be smaller,
$i.e.$ $O(g^2a)$.  In particular the variation with $\beta$ is similar
to that for Wilson fermions, however it should be noted that this is
based on data at only three values of $\beta$.  At present, looking at 
the trend in the data, the best we can do is to assume a linear behavior.

The bottom line is that the results, in the $a=0$ limit, from all
three formulations turn out to be in surprisingly good agreement with
each other.  Thus, for our best estimate of the quenched value we take
an average of the $\beta \ge 6.0$ numbers
\begin{equation}
\mbar(\MSbar,2 \GeV) = 3.4 \pm 0.4 \pm 0.3  \MeV \qquad {\rm (quenched)} .
\end{equation}
where the first error estimate is the largest of the extrapolation
errors and covers the spread in the data. The second error is a $10\%$
uncertainty due to the lattice scale as discussed in
Section~\ref{s_syserror}.

\begin{table} 
\caption{The results for the $n_f=2$ dynamical simulations listed 
in Table~\protect\ref{t_Dlist}. The notation is same as in 
Table~\protect\ref{t_mqresultsQ}.}
$$
\setlength{\tabcolsep}{1pt}
\begin{tabular}{|l|cccc|cc|cc|}
 \hline
 &&&&&\multispan2{\hfil$\mbar$\hfil\vrule}&\multispan2{\hfil$m_s(M_\phi)$\hfil\vrule}\\
 Label&%
  $\kappa_c/A_\pi$&$B_\pi$&$A_\rho$&$B_\rho$&%
  $\bar m a \times 10^5$&$\bar m$&%
  ${m_s} a  \times 10^3$&$ {m_s}$\\
  \hline

$\CW_2 1  $&$0.167928(  07)$&$ 12.471( 365)$&$-41.186(4356)$&$ 13.967(1455)$&$ 42( 05)$&$ 1.06( 06)$&$ 10(  02)$&$  25(  3)$\\
$\CW_2 2  $&$0.164866( 288)$&$  6.287( 440)$&$-10.566(1450)$&$  3.655( 466)$&$139( 20)$&$ 2.53( 20)$&$ 48(  09)$&$  88( 11)$\\
$\CW_2 3  $&$0.161335(  64)$&$  4.979( 225)$&$-11.542(1239)$&$  3.860( 393)$&$115( 11)$&$ 2.64( 15)$&$ 37(  06)$&$  85(  9)$\\
$\CW_2 4  $&$0.161218(  08)$&$  4.499( 134)$&$-14.173(1801)$&$  4.683( 576)$&$ 88( 07)$&$ 2.52( 11)$&$ 25(  04)$&$  73(  9)$\\
$\CW_2 5  $&$0.158393(  47)$&$  4.426( 285)$&$-14.249(2216)$&$  4.617( 694)$&$ 78( 13)$&$ 2.37( 21)$&$ 24(  06)$&$  73( 11)$\\
$\CW_2 6^*$&$0.158506(  13)$&$  3.903(  87)$&$-13.211( 823)$&$  4.290( 258)$&$ 85( 05)$&$ 2.65( 09)$&$ 25(  02)$&$  79(  5)$\\
\hline		  			                		
$\CS_2 1^*$&$-0.00025(   1)$&$  6.951(  66)$&$  0.423(  17)$&$  8.667( 777)$&$ 84( 07)$&$ 2.93( 12)$&$ 17(  02)$&$  58(  5)$\\
$\CS_2 2  $&$ 0.00012(   7)$&$  6.041(  98)$&$  0.323(  11)$&$  9.120( 501)$&$ 57( 04)$&$ 2.62( 09)$&$ 12(  01)$&$  57(  3)$\\
$\CS_2 3  $&$-0.00078(  11)$&$  5.573( 162)$&$  0.345(  14)$&$  7.320( 762)$&$ 67( 06)$&$ 2.95( 14)$&$ 15(  02)$&$  66(  8)$\\
\hline
\end{tabular}
$$

\label{t_mqresultsD}
\end{table}

\begin{figure}[t]
\hbox{\hskip15bp\epsfxsize=0.9\hsize \epsfbox {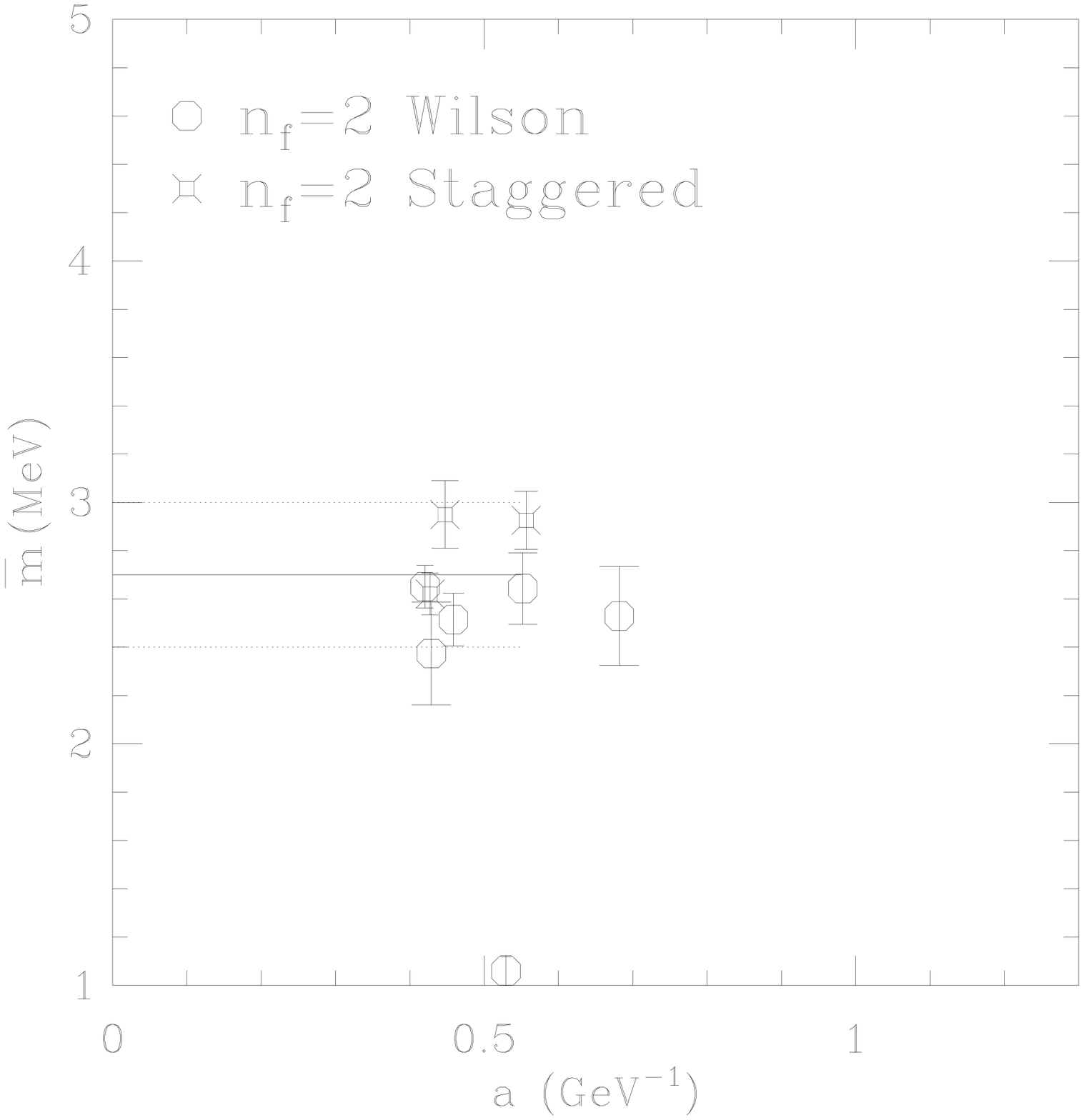}}
\vskip \baselineskip
\figcaption{The behavior of $\mbar(\MSbar,2 \GeV)$, extracted using
$M_\pi$ data for $n_f =2$ simulations. The scale is set by $M_\rho$, 
and the lattice scheme is TAD1.}
\label{f_mbarD}
\end{figure}

To analyze the $n_f =2$ dynamical configurations, we have restricted
ourselves to data with $m_{valence} = m_{sea}$.  The main limitation
we face is that the data have been obtained at very few values of
lattice spacing, and the statistics and lattice volumes are smaller
than in quenched simulations. The pattern of $O(a)$ corrections in the
present unquenched data, shown in Fig.~\ref{f_mbarD} and
Table~\ref{t_mqresultsD}, is not clear. In fact, as discussed in
Section~\ref{s_syserror}, we cannot even ascertain whether the
convergence in $a$ is analogous to the quenched case, $i.e.$ from
above for both Wilson and staggered formulations.  The strongest
statement we can make is qualitative; at any given value of the
lattice spacing, the $n_f=2$ data lies below the quenched
result. Taking the existing data at face value, we find that the
average of the Wilson and staggered values are the same for the
choices $\beta \ge 5.4$, $\beta \ge 5.5$, or $\beta \ge 5.6$.  Since
for $\beta \ge 5.5$ there are two independent measurements that agree,
our current estimate is taken to be the average of these data
\begin{equation}
\mbar(2 \GeV) =  2.7 \pm 0.3 \pm 0.3  \MeV \qquad (n_f=2 {\rm\ flavors}) \ ,
\end{equation}
where the first error estimate is the spread in the data. To obtain a value
for the physical case of $n_f=3$, the best we can do is to assume a
behavior linear in $n_f$.  In this case a linear extrapolation of the
$n_f=0$ and $2$ data gives
\begin{equation}
\mbar(2 \GeV) \approx 2.4  \MeV \qquad (n_f=3 {\rm\ flavors}) .
\end{equation}

It is obvious that more lattice data are needed to resolve the
behavior of the unquenched results. However, the surprise of this
analysis is that both the quenched and $n_f=2$ values are small and
lie at the very bottom of the range predicted by phenomenological
analyses \cite{gasserPR}.

\section{THE STRANGE QUARK MASS $m_s$}
\label{s_ms}

\begin{figure}[t]
\hbox{\hskip15bp\epsfxsize=0.9\hsize \epsfbox {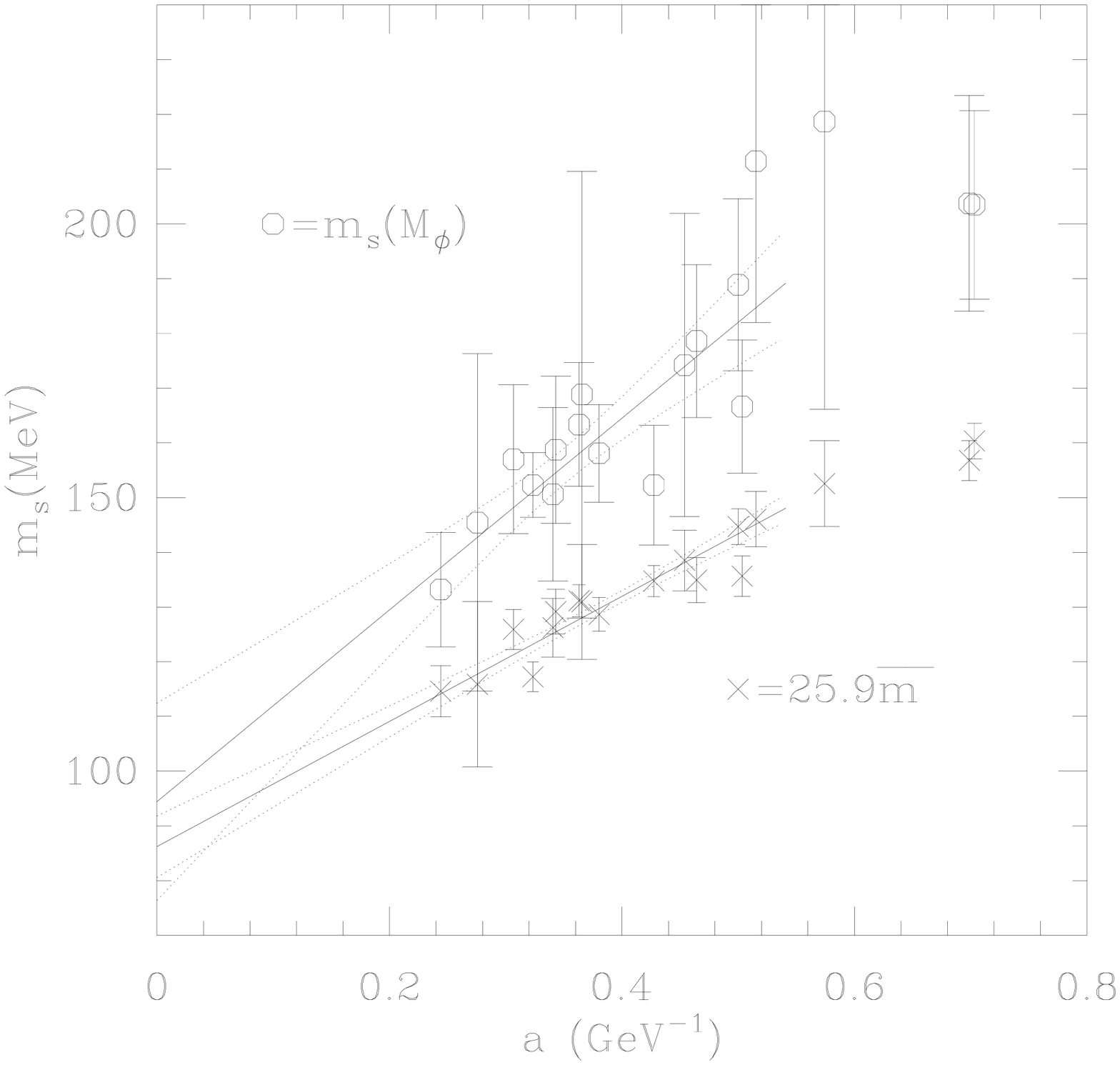}}
\vskip \baselineskip
\figcaption{Comparison of $m_s(\MSbar,2 \GeV)$ extracted using 
$M_\phi$ against the lowest order chiral prediction $m_s = 25.9\mbar$. 
The data are for the quenched Wilson theory.}
\label{f_ms}
\end{figure}

We determine $m_s$ using the three different mass-ratios, $M_K^2/M_\pi^2$, 
\ $M_{K^*} / M_\rho $, and $M_{\phi} / M_\rho$.  As mentioned above,
using a linear fit to the pseudo-scalar data constrains ${m_s(M_K) /
\mbar } = 25.9$. Unfortunately, the data are not good enough to include
terms of order $m_q^2$ or the non-analytical chiral logarithms when
making fits.  Thus we cannot improve on the lowest order results
\begin{eqnarray}
m_s(\MSbar, \mu=2 \GeV, M_{K}) &=& 88(10)  \MeV \qquad ({\rm Quenched}) \ , \nonumber \\
m_s(\MSbar, \mu=2 \GeV, M_{K}) &=& 70(8)   \MeV \qquad (n_f=2) \ ,
\end{eqnarray}
when extracting $m_s$ from $M_K^2/M_\pi^2$.  Using the
vector mesons $M_K^*$ and $ M_\phi$ gives independent estimates. To
illustrate the difference, a comparison of $m_s(M_\phi)$ and $m_s =
25.9\mbar$ is shown in Fig.~\ref{f_ms} for the quenched Wilson theory.

\begin{figure}[t]
\hbox{\hskip15bp\epsfxsize=0.9\hsize \epsfbox {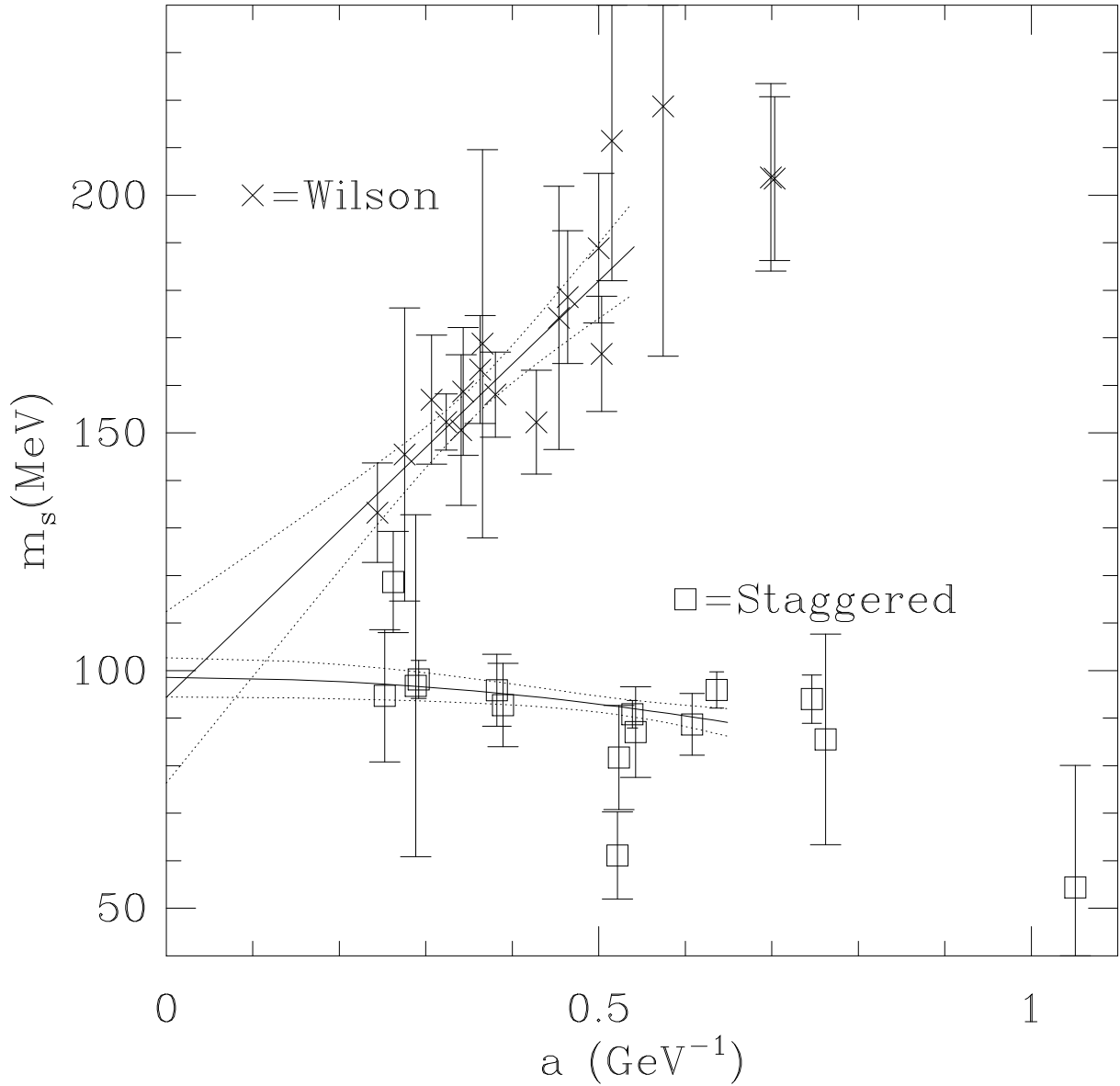}}
\vskip \baselineskip
\figcaption{Comparison of $m_s(\MSbar,2 \GeV)$ extracted using $M_\phi$ 
for the quenched Wilson, and staggered theories.}
\label{f_msphiQ}
\end{figure}

\begin{figure}[t]
\vskip \baselineskip
\figcaption{Results for $m_s(\MSbar,2 \GeV)$ extracted using $M_{K}$,
$M_{K^*}$, and $M_{\phi}$ for the quenched Clover action with
tree-level value $c_{SW}=1$ for the clover coefficient. The results
for $m_s(M_\phi)$ and $m_s(M_{K^*})$ agree.  We show linear fits to
the $m_s(M_\phi)$ and $m_s(M_{K})$ data.}
\hbox{\hskip15bp\epsfxsize=0.9\hsize \epsfbox {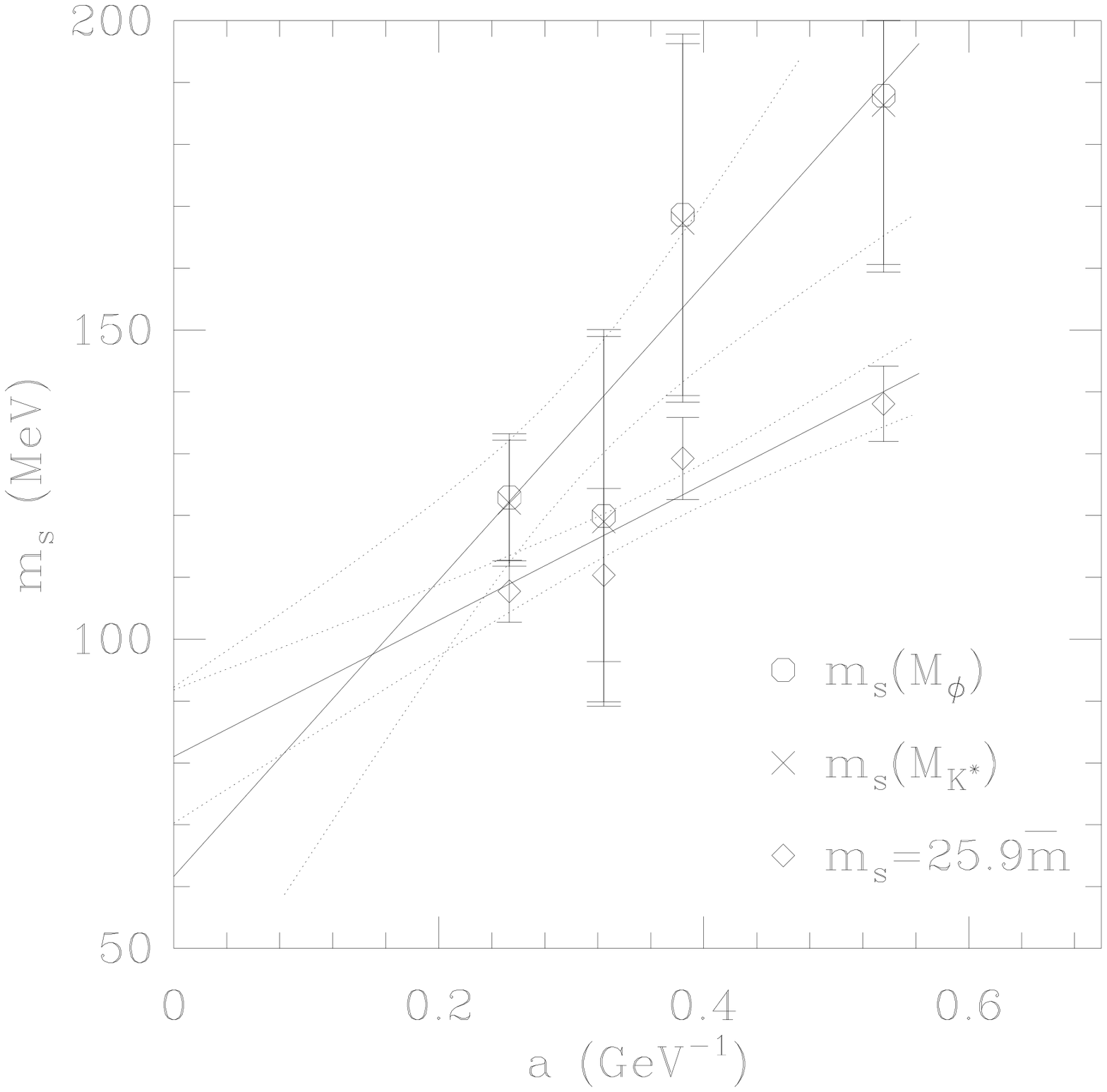}}
\label{f_clover1}
\end{figure}

The quenched Wilson and staggered data, and the extrapolation to
$a=0$, for $m_s(M_\phi)$ are shown in Fig.~\ref{f_msphiQ}. The Wilson
data again show large $O(a)$ corrections and we make a linear fit to
the data at $\beta \ge 5.93$.  The staggered data converges from
below, and anticipating $O(a^2)$ corrections we fit linearly in
$a^2$. The clover data are shown in Fig.~\ref{f_clover1}. The
parameters of the fits are
\begin{equation}
\begin{tabular}{lll}
$m_s(M_\phi) = 94(18)\MeV [1 + 1.9(7)    \GeV\ a]$    &  $\chi^2/dof=0.4$ & Wilson \\
$m_s(M_\phi) = 62(31)\MeV [1 + 3.9(2.5)  \GeV\ a]$    &  $\chi^2/dof=0.34$ & Clover \\
$m_s(M_\phi) = 99( 4)\MeV [1 - (0.48(15) \GeV\ a)^2]$ &  $\chi^2/dof=2.0$ & Staggered \,.
\end{tabular}
\label{eq:Qmsfita}
\end{equation}
Compared to $\mbar$, Wilson and clover fits have much larger errors.
This is due to the fact that the lattice measurements of the vector
mass have much larger statistical errors compared to the
pseudoscalars. Also, the spread between different groups is larger,
reflecting the differences in the strategy to extract the masses from
the 2-point correlation functions.  The fit to the staggered data for
$m_s$, on the other hand, is surprisingly much better than that for
$\mbar$.  Since the parameters of the clover fit are poorly
determined, we do not consider them any further.

If, instead, we fit the data keeping points at $ \beta \ge 6.0$, then 
\begin{equation}
\begin{tabular}{lll}
$m_s(M_\phi) = 105(21)\MeV [1 + 1.4(1.0)    \GeV\ a]$    &  $\chi^2/dof=0.3$ & Wilson \\
$m_s(M_\phi) = 104( 5)\MeV [1 - (0.74(14) \GeV\ a)^2]$ &  $\chi^2/dof=1.8$ & Staggered \,.
\end{tabular}
\label{eq:Qmsfitb}
\end{equation}
For our best estimate of $m_s(M_\phi)$ in the quenched theory 
we average the Wilson and staggered values given in Eqs.~\ref{eq:Qmsfita} and 
\ref{eq:Qmsfitb}
\begin{equation}
m_s(\MSbar, \mu=2 \GeV,M_\phi) = 100(21)  \MeV \qquad ({\rm quenched}) .
\end{equation}
where we quote the largest of the extrapolation errors. 

\begin{figure}[t]
\hbox{\hskip15bp\epsfxsize=0.9\hsize \epsfbox {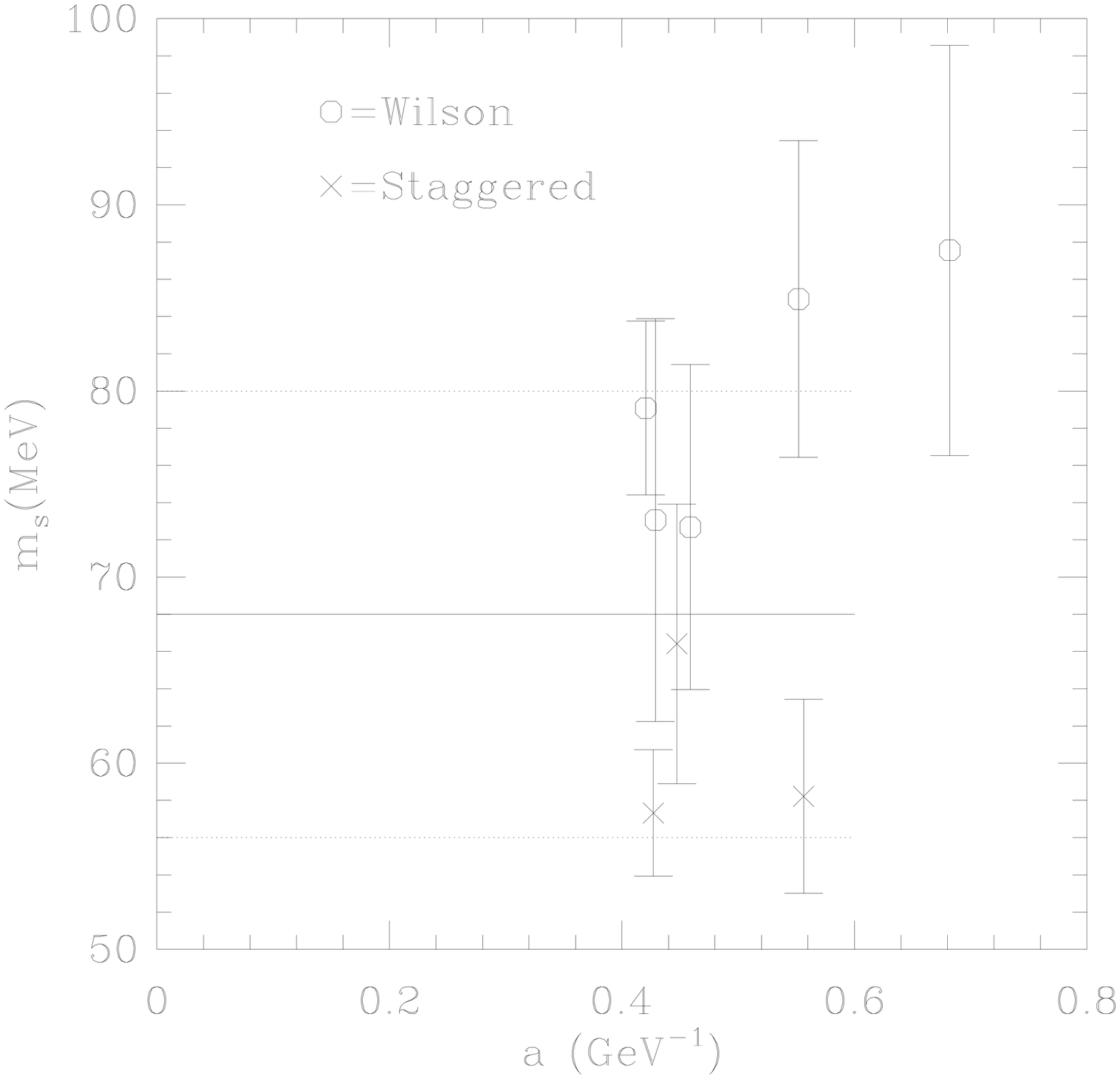}}
\vskip \baselineskip
\figcaption{Comparison of $m_s(\MSbar,2 \GeV)$ extracted using $M_\phi$ 
for the $n_f=2$ Wilson and staggered actions. The lines show the
average value and the uncertainty in the estimate for the two
formulations.}
\label{f_msphiD}
\end{figure}

The $n_f=2$ data for $m_s(M_\phi)$ are shown in Fig.~\ref{f_msphiD}.
In light of the discussion presented in Section~\ref{s_mbar} and
\ref{s_syserror} we only consider data with $\beta \ge 5.5$.
Furthermore, the spread between the values of $m_s$ obtained by
different collaborations at a given $\beta$ are large, so we do not
extrapolate assuming the expected $O(a)$ corrections. Instead we
simply take the average in both cases.  This gives $m_s = 76(10)\
\MeV$ for the Wilson formulation and $m_s = 59(6) \ \MeV$ for the
staggered.  For our final estimate in the $a=0$ limit, we take the
average of these two values,
\begin{equation}
m_s(\MSbar, \mu=2 \GeV,M_\phi) = 68(12)  \MeV \qquad (n_f=2) \ ,
\end{equation}
where the spread in the data is taken to be representative of the
uncertainty in the extrapolation to the continuum limit.

The estimates for $m_s$ found using $M_{K^*}$ match with $m_s(M_\phi)$ 
as exemplified in Fig.~\ref{f_clover1}. 
To summarize, we find that the quenched estimate of $m_s$ depends on
the hadronic state ($M_K$ versus $M_\phi$ or $M_{K^*}$) used to extract
it. While the first estimate is constrained by the lowest order chiral 
perturbation theory, the second is an independent estimate. Thus, for our 
final estimates we take 
\begin{eqnarray}
m_s(\MSbar, \mu=2 \GeV,M_\phi) &=& 100 \pm21 \pm10  \MeV \qquad ({\rm quenched}) \,, \\
m_s(\MSbar, \mu=2 \GeV,M_\phi) &=& 68 \pm12 \pm7  \MeV \qquad (n_f=2) \,.
\label{e:msfinal}
\end{eqnarray}
where the second error is the $10\%$ uncertainty in setting the
lattice scale as discussed in Section~\ref{s_syserror}. Using these,
one can again make a linear extrapolation in $n_f$ to get $m_s \sim 55
\ \MeV$ for the three flavor theory.  However, we feel that it is
important to stress that the unquenched simulations are still in very
early stages with respect to both statistical and systematic errors.
Thus, unquenched estimates should be considered preliminary.

To summarize, the data show two consistent patterns. First, for a
given value of $a$ the $n_f=2$ results are smaller than those in the
quenched approximation.  Second, the ratio $\mbar / m_s(M_\phi)$ is in
good agreement with the predictions of chiral perturbation theory for
both the $n_f=0$ and $2$ estimates.  Quantitatively, our estimates are
low and will be refined as data at more values of $\beta$, possibly
with $n_f=2-4$ to bracket the physical case of $n_f=3$, become
available.

\section{STUDY OF THE CLOVER IMPROVEMENT}
\label{s_clover}

\begin{figure}[t]
\hbox{\hskip15bp\epsfxsize=0.9\hsize \epsfbox {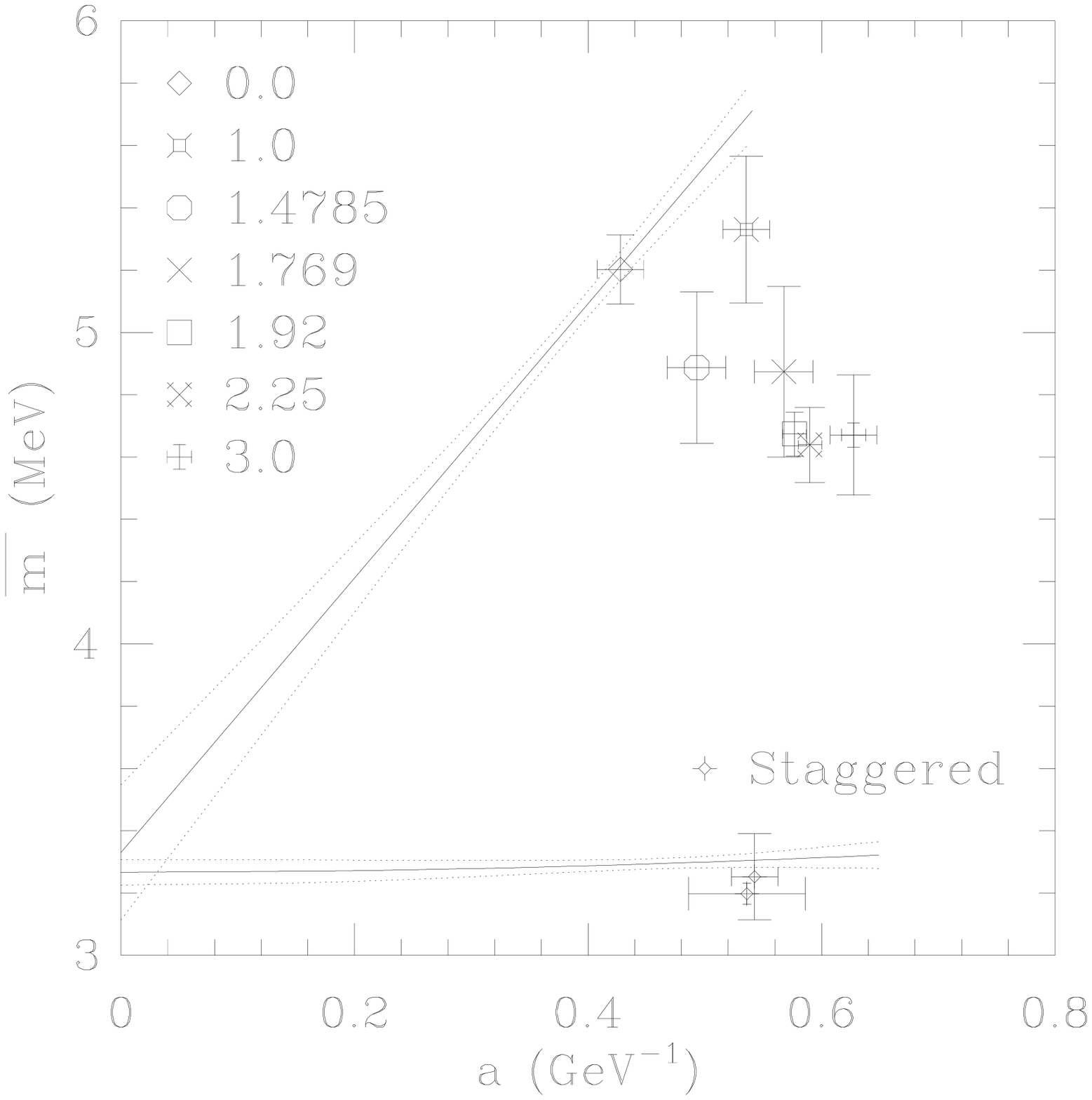}}
\vskip \baselineskip
\figcaption{Behavior of $\mbar(\MSbar,2 \GeV)$ versus scale $a$ as a function of 
the clover coefficient $c_{SW}$. We also reproduce the fits to the Wilson and staggered 
data, and the staggered result at $\beta=6.0$ from Fig.~\ref{f_mbar} for comparison. }
\label{f_cimp}
\end{figure}

At $\beta=6.0$ there now exist data at a number of values of the
clover coefficient as listed in Table~\ref{t_Clist}.  This allows us
to study the effect of improvement on quark masses as a function of
$c_{SW}$ in the range 0--3.0.  Note that the value predicted by the
Alpha Collaboration for $O(a)$ non-perturbative improvement is $1.769$
\cite{luscher96imp}.  The data, presented in Table~\ref{t_mqresultsC},
are plotted in Fig.~\ref{f_cimp} along with the Wilson and staggered
fits reproduced from Fig.~\ref{f_mbar}, and the staggered points at
$\beta=6.0$.  Qualitatively, increasing the clover coefficient
increases $a$ and decreases $\mbar$. In particular, the scale at
$c_{SW}=1.769$ matches that from the staggered data.  However, the
value of $\mbar$ is significantly ($\sim 50\%$) different.  For a
complete $O(a)$ improvement with clover actions, the lattice quark
mass needs an $O(ma)$ correction, $i.e.$ $m_q^{improved} = m_q ( 1 +
b_m m_q a)$ where $b_m = -0.5$ at tree level \cite{luscher96imp}.
This is, however, a few percent effect only for the parameter values
we have analyzed.  The other two possible explanations are the failure
of perturbation theory used to determine $Z_m$, and possibly large
$O(a^2)$ effects. More work, like determining $Z_m$
non-perturbatively, is required to quantify these effects.

\begin{table} 
\caption{Lattice parameters of the runs at $\beta=6.0$ for different values of the 
clover coefficient in the Sheikholeslami-Wohlert action.  \looseness=-1}
\vskip 12pt
\def\q{\quad}
\def\s{\phantom{-}}
\newcommand\0{hphantom{0}}
\setlength{\tabcolsep}{2pt}
\begin{tabular}{|l|c|c|c|c|l|c|cc|}
\hline
      &      &         & Lattice & \# of &                           & Scale $1/a$  &%
          \multispan{2}{\hfil$\alpha_{\MSbar}(q^*)$\hfil\vrule}\\
Label & Ref. & $c_{SW}$ & size    & Conf. & Quark masses used in the fits & (GeV)         &%
          $(1/a)$&$(2\GeV)$\\
\hline
W8 & \cite{rHM96LANL}    & 0       &$32^3\times 64$& 170& 0.155,   0.1558,  0.1563          &$2.338( 43)$&$0.192$&$0.205$\\
C1 & \cite{APE96}        & 1.0     &$18^3\times 64$& 200& 0.1425,  0.1432,  0.1440          &$1.867( 81)$&$0.192$&$0.187$\\
C5 & \cite{rHM97LANL}    & 1.4785  &$32^3\times 64$& 100& 0.13808, 0.13851, 0.13878         &$2.028( 96)$&$0.192$&$0.193$\\
C6 & \cite{HM96SCHIER}   & 1.769   &$24^3\times 64$& 100& 0.1342,  0.1346,  0.1348          &$1.762( 78)$&$0.192$&$0.184$\\
C7 & \cite{HM96SCHIER}   & 1.92    &$16^3\times 32$& 100& 0.1290,  0.1300,  0.1310, 0.1320  &$1.734( 26)$&$0.192$&$0.182$\\
C8 & \cite{HM96SCHIER}   & 2.25    &$16^3\times 32$& 100& 0.1260,  0.1265,  0.1270, 0.1277  &$1.695( 42)$&$0.192$&$0.181$\\
C9 & \cite{HM96SCHIER}   & 3.0     &$16^3\times 32$& 100& 0.1160,  0.1165,  0.1170, 0.1173  &$1.594( 48)$&$0.192$&$0.177$\\
\hline
\end{tabular}

\label{t_Clist}
\end{table}

\begin{table} 
\caption{The results for the quenched simulations using the
Sheikholeslami-Wohlert action as listed in
Table~\protect\ref{t_Clist}. The notation is the same as in
Table~\protect\ref{t_mqresultsQ}.}
$$
\setlength{\tabcolsep}{1pt}
\begin{tabular}{|l|cccc|cc|cc|}
 \hline
 &&&&&\multispan2{\hfil$\mbar$\hfil\vrule}&\multispan2{\hfil$m_s(M_\phi)$\hfil\vrule}\\
 Label&%
  $\kappa_c   $&$B_\pi$&$A_\rho$&$B_\rho$&%
  $\bar m a \times 10^5$&$\bar m$&%
  ${m_s} a  \times 10^3$&$ {m_s}$\\
  \hline
W$ 8$&$0.157143(    03)$&$  1.993(  26)$&$ -6.663( 521)$&$  2.196( 162)$&$172( 06)$&$ 5.20( 11)$&$50(  05)$&$ 152(  11)$\\
C$ 1$&$0.145483(    03)$&$  2.674(  29)$&$ -6.265(1006)$&$  1.942( 288)$&$201( 18)$&$ 5.33( 24)$&$71(  13)$&$ 188(  27)$\\
C$ 5$&$0.139285(    01)$&$  2.837(  50)$&$ -9.913(2491)$&$  2.866( 689)$&$161( 15)$&$ 4.89( 24)$&$45(  12)$&$ 136(  32)$\\
C$ 6$&$0.135222(    05)$&$  3.268( 126)$&$ -7.318(2973)$&$  2.096( 799)$&$185( 17)$&$ 4.87( 27)$&$70(  29)$&$ 183(  68)$\\
C$ 7$&$0.133062(    05)$&$  3.483(  20)$&$ -8.914( 243)$&$  2.489(  63)$&$179( 05)$&$ 4.67( 07)$&$60(  02)$&$ 156(   4)$\\
C$ 8$&$0.128536(    07)$&$  3.640(  50)$&$ -9.934( 673)$&$  2.669( 170)$&$179( 09)$&$ 4.64( 12)$&$57(  05)$&$ 147(   9)$\\
C$ 9$&$0.118151(   104)$&$  3.906( 195)$&$-13.101( 714)$&$  3.209( 166)$&$189( 10)$&$ 4.67( 19)$&$51(  04)$&$ 125(   6)$\\
\hline
\end{tabular}
$$

\label{t_mqresultsC}
\end{table}

\section{ANALYSIS OF SYSTEMATIC ERRORS}
\label{s_syserror}

We have checked the validity of our analysis ignoring correlations
between data by comparing results presented here against those
obtained from a detailed error analysis using the quenched Wilson data
obtained on $170$ $32^3 \times 64$ lattices at $\beta=6.0$
\cite{rHM96LANL}.  For example, in Ref.~\cite{rHM96LANL}, we quote
$m_s(M_\phi) =154(8)\ MeV$, whereas the current analysis gives
$152(11)$, the number plotted in Fig.~\ref{f_msphiQ}.  A similar
agreement is seen if we compare our estimates with the results of the
original analysis by collaborations whose data we have used as
discussed in Section~\ref{s_comparison}.  Thus, we believe that the
analysis presented here gives central values in the individual points
reliable to within a few percent, and the estimate of statistical
errors to within a factor of two.

There is accumulating evidence in high statistics data that fits
linear in the quark mass, Eq.~\ref{e:chiralfits}, are not sufficient,
$i.e.$ chiral logs and terms of higher order in $m$ need to be
included \cite{mq94gupta, REV96SHARPE, LAT96gottlieb}.  (Chiral
perturbation theory for pseudoscalar mesons does not indicate any large 
non-linearities as the change in $m_s/\mbar$ and $m_u/m_d$ between 
the lowest and next order analyses is insignificant \cite{rMq96leutwyler}.)
In a state-of-the-art quenched calculation presented in
Ref.~\cite{rHM96LANL}, we found that including higher order terms in
the fits for pseudoscalar and vector mesons changes the estimates of
quark masses by less than a few percent.  In the present study we have
not carried out this detailed analysis because in most of the data the
quality of statistical precision and the number of values of light
quark masses simulated are insufficient to uncover these effects.  The
deviations from linearity are found to be smaller for pseudoscalar
mesons, which is another reason why we used them to fix $\mbar$, and
vector mesons to fix $m_s$. In short, even though present lattice data
does not rule out the possibility of large systematic errors due to
the neglected higher order terms, there is no clear indication
that this is the case. This issue can be addressed only when the range
of quark masses used in the fits is increased.

For Wilson and clover type of fermions, the errors in the calculation
of $\kappa_c$ and the $\kappa$ corresponding to light quarks is highly
correlated. As a result the estimate of errors in $1/2\kappa -
1/2\kappa_c$ is much smaller, and we find that the low statistics
points give estimates that are consistent with the high statistics
large lattice data. Thus we believe that the estimate, in the $a=0$
limit, will not change significantly on improving just the statistical
quality of the current data.  What is more important is to increase 
the range of quark masses in the chiral fits and to verify
whether the lowest order correction formulae used in
the extrapolation to $a=0$ are sufficient. We have partially addressed
the second issue by comparing the continuum limit of data from different
lattice actions. We plan to test these issues in the future as data
at smaller $m_q$ and $a$ become available.

The clover action with $c_{SW}=1.0$ has discretization errors starting
at $O(\alpha_s a)$ and it is not known whether $O(\alpha_s a)$ or
$O(a^2)$ errors dominate for $\beta \ge 6.0$.  Nevertheless, one would
have expected the corrections to be smaller than those for Wilson
fermions.  However, the current data show a near agreement between
Wilson and clover data. With data at just three values of coupling it
is hard to determine what form to use to extrapolate the clover action
data to the continuum limit.  The most we can say is that using a
linear extrapolation, a theoretically allowed form, gives results
consistent with the rest of our analysis.

\noindent{\bf The lattice scale \protect\boldmath $a$}: The value of
the lattice scale $a$ enters into the calculation in two ways. It is
used to convert the dimensionless lattice quantity $ma$ into physical
units. A much weaker dependence arises when matching the lattice and
continuum theories.  This is because one specifies the renormalization
point $\mu$, at which the lattice and continuum theories are matched,
in terms of $1/a$. Present lattice data show a considerable spread in
the determination of $a$ depending on the physical quantity used. This
variation is due to a combination of discretization and quenching
errors.  For example, with our high statistics quenched Wilson
data~\cite{rHM96LANL}, we have calculated the lattice scale fixing
$M_\rho$, $f_\pi$, and $M_N$ to their physical values. The results are
$1/a(M_\rho) = 2.330(41) \GeV$, $1/a(f_\pi) = 2.265(57) \GeV$, and
$1/a(M_N) = 2.062(56) \GeV$ \cite{rHM96LANL}.  Also, the fluctuations
between the determination of scale at a given $\beta$ by the different
groups (see Tables~\ref{t_Qlist}, \ref{t_Dlist}, \ref{t_Clist}) are of
the same magnitude.  Thus, using $M_\rho$ to set the scale could lead
to an additional $10\%$ uncertainty in our estimates of quark masses,
which we quote as a separate error in our final estimates.

\begin{figure}[t]
\hbox{\hskip15bp\epsfxsize=0.9\hsize \epsfbox{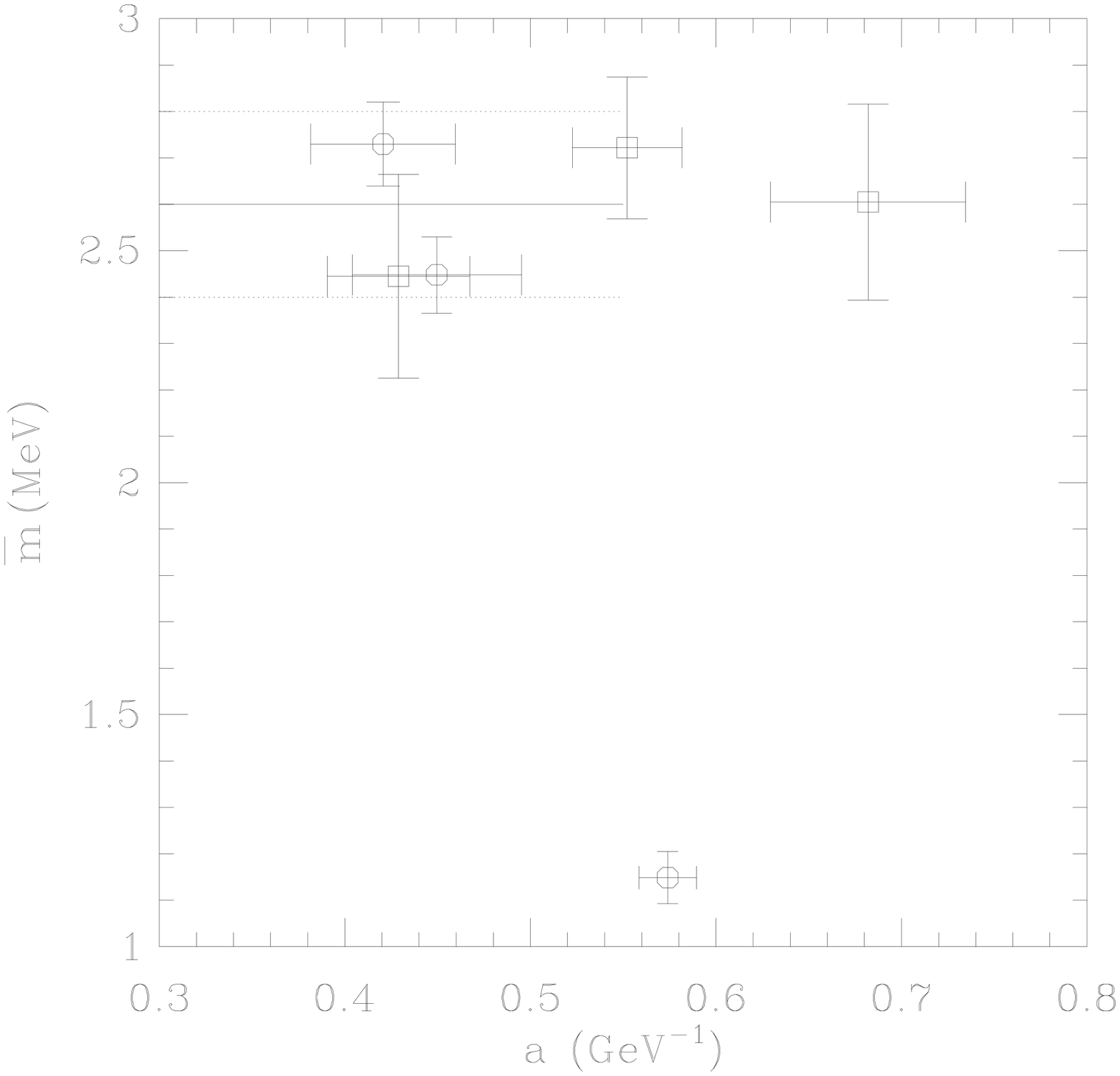}}
\vskip \baselineskip
\figcaption{The $n_f=2$ data for Wilson fermions. We show the highest
statistics points at $\beta=5.3, 5.5, 5.6$ with the symbol
octagon. The remaining points are shown by the symbol square. We also
plot our estimate, the average value for $\beta \ge 5.5$, and show the
errors in $a$ for each data point.}
\label{f_martifact}
\end{figure}

\noindent{\bf The \protect\boldmath $n_f=2$ Wilson data}: We show the
data for $\beta \ge 5.3$ obtained on $16^3 \times 32$ lattices in
Fig.~\ref{f_martifact}. We highlight the highest statistics points by
the symbol octagon. The striking feature of this data is the
surprisingly low value $\mbar\approx 1.1\MeV$ at
$\beta=5.3$. Coincidentally, the SCRI collaboration has reported an
anomalous behavior in the calculation of the non-perturbative
$\beta$-function at $\beta=5.3$ \cite{dbetaW296SCRI}. The important
question is whether this is an anomalous measurement or is 
indicative of a serious problem.  If this feature is borne out by
future calculations, then one possible interpretation is that the
lattice theory has a phase transition close to $\beta=5.3$, similar to
that seen in the fundamental-adjoint coupling plane of the pure gauge
model.  Such singular points are lattice artifacts, and to avoid their
influence in the extrapolation to $a=0$ one has to consider data at
weaker coupling or improve the action to stay away from such
singularities. In the former case one would need to consider data at
couplings weaker than $\beta = 5.3$. Motivated by the need to minimize
contamination from such effects, we have chosen to use data at $\beta
\ge 5.5$ only.

The above scenario would also explain why the Wilson data for $\beta
\le 5.5$ ($a > 0.42 \GeV^{-1}$) appear to lie marginally below the
staggered data, in contrast to the quenched estimates.  Furthermore,
one might worry about the presence of such artifacts at even weaker
coupling. In the long run such a possibility is addressed by our
strategy of demanding agreement, in the $a=0$ limit, of results
obtained using different lattice discretizations of the Dirac action
since the presence and influence of such artifacts depend on the
lattice action.

\section{COMPARISON WITH PREVIOUS ESTIMATES}
\label{s_comparison}

The ROME/APE collaboration quotes the value $m_s(M_K) =128(18)\ MeV$
\cite{Allton94} from their quenched simulations. Their estimate is
based on an analysis of data obtained using both the Wilson and the
clover actions at $\beta=6.0$ and $6.2$, but
does not include a linear extrapolation to $a=0$.  Their number is
consistent with the result $128(4) \MeV$ reported by us in
Ref.~\cite{rHM96LANL}.  However, our present analysis of the global
data show the presence of large $O(a)$ corrections for the Wilson
theory.  Including this correction results in a much lower number,
$m_s(M_K)=88(10)\ MeV$, after extrapolation to $a=0$ as shown in
Fig.~\ref{f_ms}.

The APE collaboration has recently updated their estimates using data
with Wilson and Clover actions at $\beta=6.0,\ 6.2,\ 6.4$
\cite{APE96}. These data are included in our analysis and we agree with
the raw numbers.  However, they quote the value $m_s(\MSbar, 2\ \GeV,
M_K) = 122(20) \MeV$, based on taking an average of the data rather
than doing an extrapolation in $a$. Since they only analyze their own
data, they are not able to make a case for linear extrapolation as
advocated by us.  

Gough \etal\ \cite{MQ96GOUGH} have also presented results for both the
Wilson and clover actions for $\beta=5.7,\ 5.9,\ 6.1$. Even though
their values for $\mbar(\MSbar, 2 \GeV)$ at a given $\beta$ are
systematically lower than what we find due to the differences between
the two analyses in determining $a$, their estimates of the continuum
values are consistent with the results presented here.

The analysis by Lee of the $n_f=2$ staggered data generated by the
Columbia collaboration at $\beta=5.7$ yielded $\mbar = 2.7 \MeV$
\cite{rmq93Lee}. This is consistent with $2.62(9)$, the result of our
analysis of their data, and with the value, $2.95(14) \MeV$, we get
for the data by Fukugita \etal\ \cite{Tsukuba94hmks2} at the same
coupling.

To conclude, the values of quark masses that we have extracted at a
given lattice scale are consistent with previous analysis.  Both the
central value and the error estimates based on our simple analysis are
consistent with the previous analysis once the differences in the
regularization schemes are taken into account.  The main new feature
of our analysis is to show that the data roughly follow the expected
$O(a)$ corrections for quenched Wilson and clover fermions and
$O(a^2)$ for staggered. We also show that including these corrections
give results that are consistent between the various discretization
schemes.

\section{IMPACT ON $\epsilon'/\epsilon$}
\label{s_epsilon}

The Standard Model (SM) prediction 
of $\epsilon'/\epsilon$ can be written as \cite{rCP96Buras}
\begin{equation} 
\epsilon'/\epsilon = A \bigg\{c_0 + \big[c_6 B_6^{1/2} + c_8 B_8^{3/2} \big] M_r \bigg\} \ ,
\end{equation}
where $M_r = (158\MeV/(m_s + m_d))^2$ and all quantities are to be
evaluated at the scale $m_c = 1.3 \GeV$. This form highlights the
dependence on the light quark masses and the $B$ parameters.  For
central values of the SM parameters quoted by Buras \etal\
\cite{rCP96Buras}, we estimate $A = 1.29\times 10^{-4}$, $c_0 = -
1.4$, $c_6 = 7.9$, $c_8 = - 4.0$. Thus, to a good approximation
$\epsilon'/\epsilon \propto M_r$; and increases as $B_8^{3/2}$
decreases.  Conventional analysis, with $m_s + m_d = 158 \MeV$ and
$B_6^{1/2} = B_8^{3/2} = 1$, gives $\epsilon'/\epsilon \approx 3.2
\times 10^{-4}$.  On the other hand taking $m_s + m_d \approx 85
\MeV$, our $n_f=2$ estimates scaled to $m_c$, and $B_8^{3/2} = 0.8$
\cite{BKW96LANL} gives $\epsilon'/\epsilon \approx 19 \times 10^{-4}$.
This estimate lies in between the Fermilab E731
($7.4(5.9)\times10^{-4}$) \cite{epsE731} and CERN NA31
($23(7)\times10^{-4}$) \cite{epsNA31} measurements. Since the new
generation of experiments will reduce the uncertainty to $1
\times10^{-4}$, tests of the enhanced value are imminent.

\section{Continuum limit of the Quark Condensate}
\label{s_xxgmor}

\begin{figure}
\hbox{\hskip15bp\epsfxsize=0.9\hsize\epsfbox {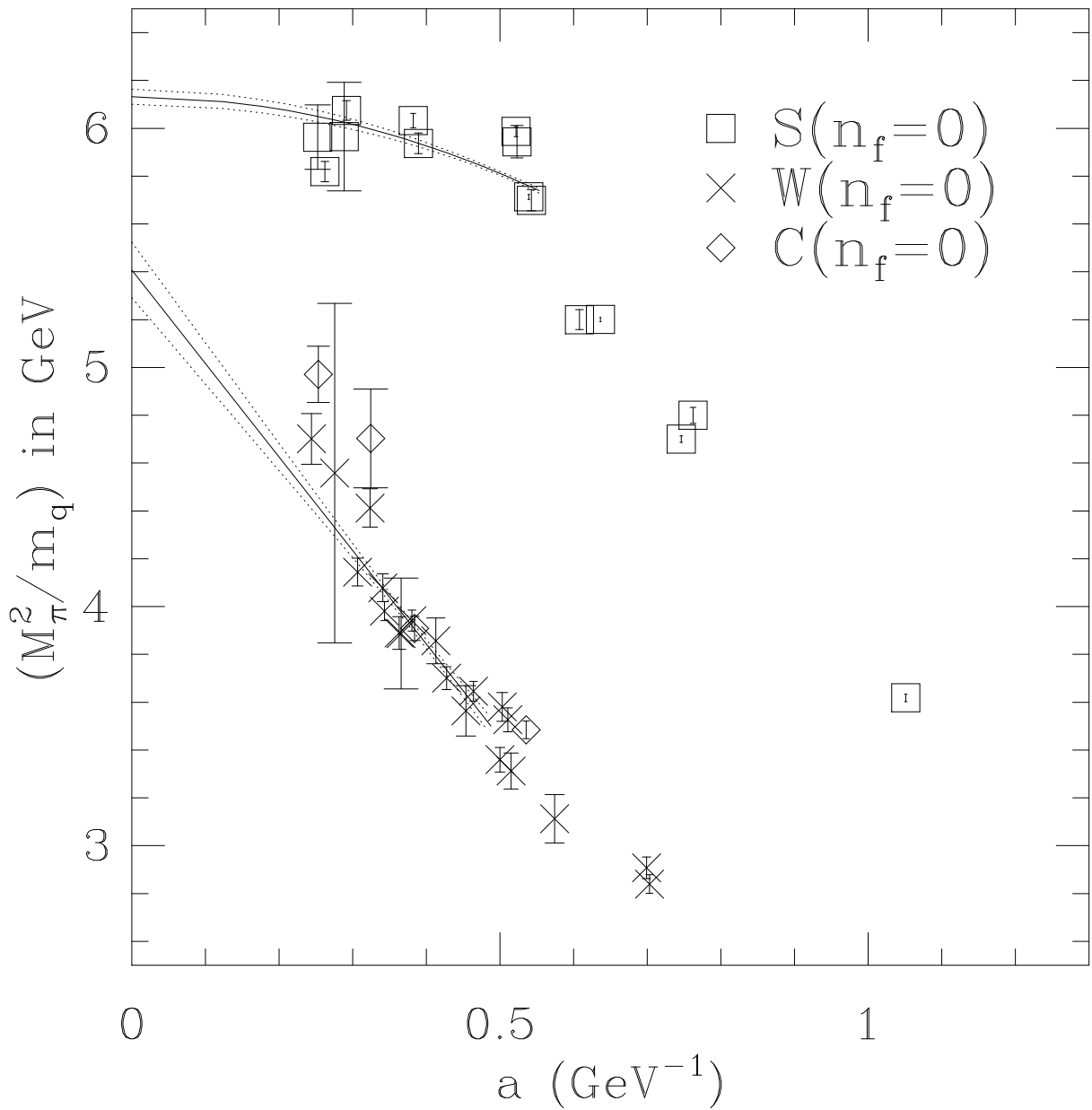}}
\vskip \baselineskip
\figcaption{The slope of $M_\pi^2$ versus $m_q(\MSbar, 2 \GeV)$ for
the quenched data as a function of the lattice spacing $a
(M_\rho)$. For clarity, the fit to the clover data is not shown.}
\label{f_Qslopes}
\end{figure}

\begin{figure}
\hbox{\hskip15bp\epsfxsize=0.9\hsize\epsfbox {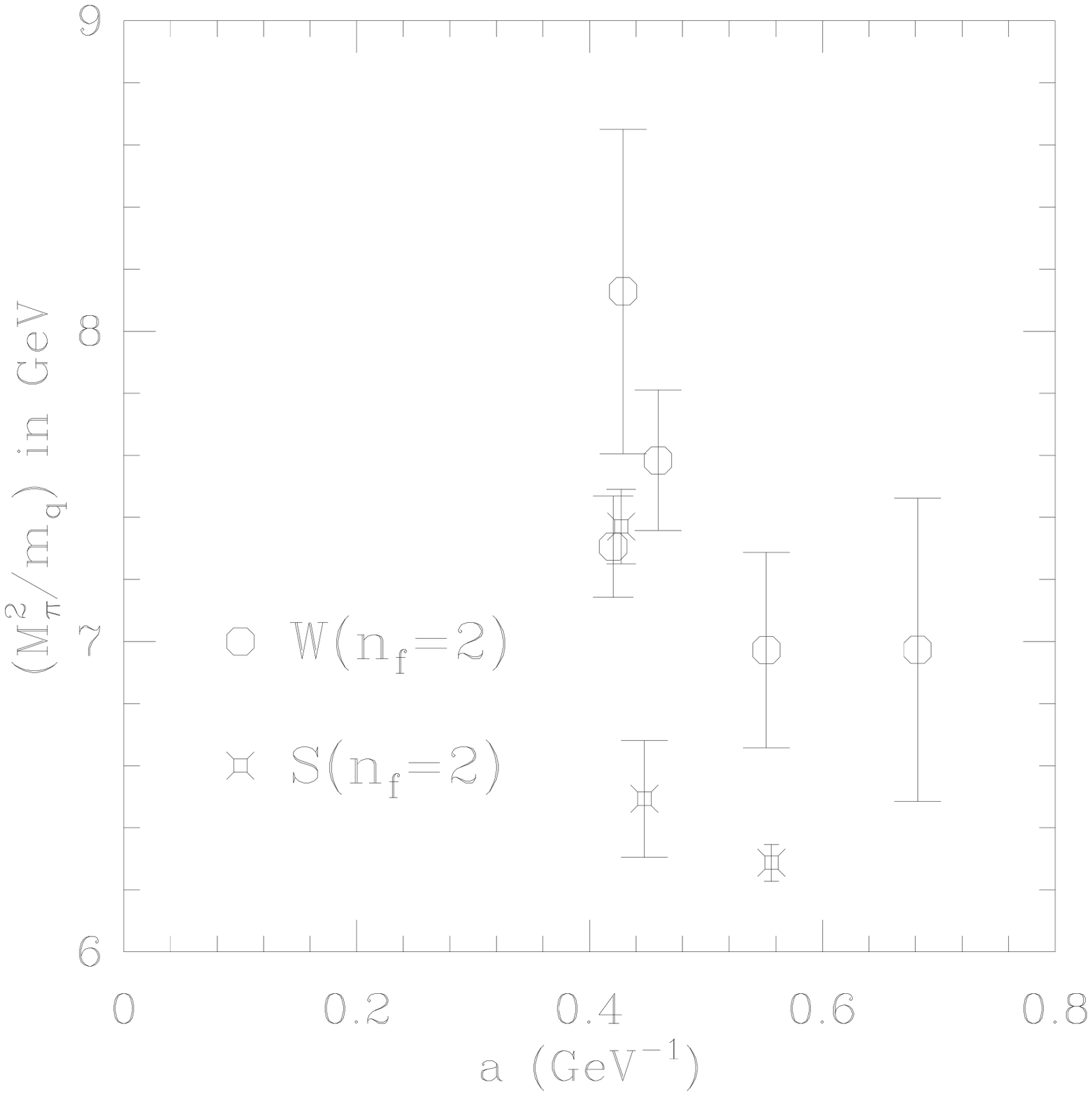}}
\vskip \baselineskip
\figcaption{The slope of $M_\pi^2$ versus $m_q(\MSbar, 2 \GeV)$ for
the dynamical data as a function of the lattice spacing $a (M_\rho)$.}
\label{f_Dslopes}
\end{figure}

The observation that the present lattice data for pseudoscalar mesons
is well described by $M_\pi^2 = B_\pi m_q$ allows us to calculate the
chiral condensate using the Gell-Mann-Oakes-Renner
relation~\cite{gasserPR, rGMOR}
\begin{equation}
\vev{\bar\psi\psi}^{\rm GMOR}
	= \lim_{m_q\to0} \ -{ f_\pi^2 M_\pi^2 \over {4 m_q}}.
\label{eq:GMOR}
\end{equation}
Since $m_q \vev{\bar\psi\psi} $ is renormalization group invariant, we
analyze the slope of $M_\pi^2$ versus the renormalized mass at a fixed
scale, $i.e.$ $m_q(\MSbar, 2\GeV)$, for the data sets described in
Tables~\ref{t_Qlist} and \ref{t_Dlist}. The slope, after extrapolation to
$a=0$, directly give an estimate of $4
\vev{\bar\psi\psi}/f_\pi^2$ in $\MSbar$ scheme at $\mu=2\GeV$. Our
results for the quenched and dynamical configurations are displayed in
Figs.~\ref{f_Qslopes} and \ref{f_Dslopes} respectively.  The fits, 
assuming a linear behavior in $a$ for Wilson and clover and $a^2$
for staggered fermion formulations, to the quenched data with 
$\beta \ge 6.0$ give
\begin{eqnarray}
\frac{-4 \vev{\bar\psi\psi}}{f_\pi^2} = 5.41(12) \GeV [1 -  0.72(7) \GeV\ a]&\chi^2/dof = 2.6\ (Wilson\ n_f=0)\\
\frac{-4 \vev{\bar\psi\psi}}{f_\pi^2} = 5.64(15) \GeV [1 -  0.73(6) \GeV\ a]&\chi^2/dof = 12.1 (clover\ n_f=0)\\
\frac{-4 \vev{\bar\psi\psi}}{f_\pi^2} = 6.13( 3) \GeV [1 - (0.46(2) \GeV\ a)^2]&\chi^2/dof = 19.5 (staggered\ n_f=0)
\end{eqnarray}
The fits for clover and staggered data do not work. In particular, the
staggered data show a break at $\beta \approx 6.0$. At weaker coupling
the data lie in the band $5.9(2)$ showing no clear $a$ dependence.
Taking this to be the best estimate for staggered fermions, we find
that the three values are in rough agreement. Thus, for our final
value we take the mean of these, $i.e.$ $4 \vev{\bar\psi\psi}/f_\pi^2
= 5.7(4) \GeV$, where the error covers the spread. Using the
experimental value of $f_\pi = 131 MeV$, we then obtain
\begin{eqnarray}
    \vev{\bar\psi\psi} &{}=  -0.024  \pm 0.002  \pm 0.002  \GeV^3 \\
m_s \vev{\bar\psi\psi} &{}=  -0.0023 \pm 0.0006 \pm 0.0004 \GeV^4
\end{eqnarray}
as our estimate for the quenched theory. The second error arises from
the 10\% scale uncertainty discussed in section \ref{s_syserror}.

The behavior of the dynamical data is again not clear. Therefore, we
take the average $7.3(1.0)$ of the values at the weakest coupling as
our best estimate.  This gives
\begin{eqnarray}
    \vev{\bar\psi\psi} &{}=  -0.031  \pm 0.004  \GeV^3 \\
m_s \vev{\bar\psi\psi} &{}=  -0.0021 \pm 0.0006 \GeV^4 \,.
\end{eqnarray}
The corresponding phenomenological estimates are \cite{gasserPR}
\begin{eqnarray}
    \vev{\bar\psi\psi} &{}=  -0.0114 \GeV^3 \\
m_s \vev{\bar\psi\psi} &{}=  -0.0017 \GeV^4 \,.
\end{eqnarray}
Thus, lattice estimates give low values for the quark masses and correspondingly 
high values for the condensate, while roughly preserving $m \vev{\bar\psi\psi} $. 

\section{CONCLUSIONS:} 
\label{s_conclusions}

We have presented an analysis of $\mbar$ and $m_s$ using data
generated by us over the years and also by other collaborations. The
values of quark masses we have extracted for a given set of lattice
parameters are consistent with previous analyses.  Both the central
value and the error estimates we get by reanalyzing the data are
consistent with the previous reported results once the differences in
the regularization schemes are taken into account.  The main new
feature of our work, based on the global analysis, is to show that the
data roughly follow the expected $O(a)$ corrections for Wilson
fermions and $O(a^2)$ for staggered.  The errors in the clover action
are expected to be $O(a/log(a))$, however over the limited range of
$a$ it is not surprising that the present data is consistent with a
linear fit.  What is surprising is that the $O(a)$ corrections are
still as large as for Wilson fermions. The bottom line is that
including these corrections give results that are consistent between
the three discretization schemes after extrapolation to the continuum
limit.

We quote our final results in the $\MSbar$ scheme evaluated at $\mu =
2 \GeV$.  The lattice perturbation theory is reorganized using the
Lepage-Mackenzie scheme.  Our best estimate of the isospin symmetric
mass $\mbar$ is $ 3.4 \pm 0.4 \pm 0.3 \MeV$ for the quenched
theory. For the $n_f=2$ flavors there does not exist enough data to
extrapolate to the continuum limit. The mean of the data at weakest
couplings gives $ \mbar(\MSbar, 2\ \GeV) = 2.7 \pm 0.3 \pm 0.3 \MeV$.
Using a linear extrapolation in $n_f$ would give $\mbar \sim 2.4 \MeV$
for the physical case.

To extract the value of $m_s$ we use the physical value of $M_\phi$
(or equivalently $M_{K^*}$). Using $M_K$ instead constrains $m_s(M_K) =
25.9 \mbar$ since we use a linear fit to the pseudoscalar mass
data. Our best estimate for the quenched theory is $m_s(\MSbar, \mu=2
\GeV, M_\phi) = 100 \pm 21 \pm 10 \MeV$, and $68 \pm 12 \pm 7 \ \MeV$
for the two flavor case. The variation with $n_f$ would again suggest
an even smaller value for the physical $n_f = 3$ theory, however, we
caution the reader that the unquenched simulations are still in infancy. 

In short, taking into account the various systematic errors, like 
allowing for a $10\%$ uncertainty in the estimates due to
the uncertainty in setting the lattice scale $a$, the present lattice
results for both $\mbar$ and $m_s$ are surprisingly low compared to
the numbers used in phenomenology, $i.e.$ $\mbar = 6-7 \MeV$ and $m_s
= 150-175\ \MeV$. The main uncertainty in the lattice results arises
due to lack of control over the extrapolation of the two flavor data
to the continuum limit, and consequently the final extrapolation to
$n_f=3$. To address these issues requires unquenched data at more
values of $\beta$.  The data suggest that, with respect to statistical
and discretization errors, the better approach is to use staggered
fermions, however one has to confront the issue of a large $Z_m$. On
the other hand one needs to understand why the non-perturbatively
improved Sheikholeslami-Wohlert action has large discretization
errors.  Clearly, the reliablity of these estimates will be improved
in the next few years as more data become available.

From a study of the variation of the pseudoscalar data as a function
of the quark mass we also extract the quantity $\vev{\bar \psi \psi} /
f_\pi^2$ using the Gell-Mann-Oakes-Renner relation. After taking the
continuum limit we find that the chiral condensate is roughly a factor
of two larger than phenomenological estimates. Consequently, the
estimate of the renormalization group invariant quantity $ m \vev{\bar
\psi \psi}$ is preserved.

\section{Acknowledgements}

The simulations carried out by our collaboration have been done on the
Crays at NERSC, and on the CM2 and CM5 at LANL as part of the DOE HPCC
Grand Challenge program, and at NCSA under a Metacenter allocation.
We thank Jeff Mandula, Larry Smarr, Andy White and the entire staff at
the various centers for their tremendous support throughout this
project. We also thank Chris Allton, Shoji Hashimoto, Urs Heller,
Thomas Lippert, Don Sinclair, and Akira Ukawa for communicating some
of their unpublished data to us.  We are grateful to Eduardo de Rafael
for discussions on the sum-rule analysis, and to Steve Sharpe for his
comments.

\end{document}